\documentclass{iopjournal}
\usepackage{amsmath}
\usepackage{ragged2e}
\usepackage{caption}

\begin{document}

\articletype{Topical Review} 

\title{Non-equilibrium Green's function formalism for radiative heat transfer}

\author{Yahan Liu$^1$\orcid{0009-0000-1900-2880} and Tao Zhu$^{2,*}$\orcid{0000-0002-0335-6657}}

\affil{$^1$School of Physical Science and Technology, Tiangong University, Tianjin, China}

\affil{$^2$School of Electronic and Information Engineering, Tiangong University, Tianjin, China}

\affil{$^*$Author to whom any correspondence should be addressed.}

\email{zhutao@tiangong.edu.cn}

\keywords{Non-equilibrium Green's function, Radiative heat transfer, Quantum transport}

\begin{justify}
\begin{abstract}
\begin{justify}
	Radiative heat transfer (RHT) at the nanoscale can vastly exceed the far-field blackbody limit due to the tunneling of evanescent waves, a phenomenon traditionally described by fluctuational electrodynamics (FE). While FE has been exceptionally successful for systems in local thermal equilibrium, its foundational assumptions break down in the growing number of scenarios involving genuine non-equilibrium conditions, such as in active devices or driven materials. This review introduces the non-equilibrium Green's function (NEGF) formalism as a powerful and versatile framework to study RHT beyond these classical limits. Rooted in quantum many-body theory, NEGF provides a unified language to describe energy transport by photons, electrons, and phonons on an equal footing. We first outline the theoretical foundations of the NEGF approach for RHT, demonstrating how it recovers the canonical results of FE in the local equilibrium limit. We then survey recent breakthroughs enabled by NEGF, including: (i) providing a quantum-accurate description of equilibrium RHT that naturally incorporates non-local and finite-size effects, resolving unphysical divergences predicted by local models; (ii) unifying heat transfer channels to reveal the non-additive synergy between radiation, electron tunneling, and phonon conduction at sub-nanometer gaps; (iii) enabling the quantum design of materials and metamaterials with tailored thermal properties through band structure and topological engineering; and (iv) describing active control of heat flow in driven systems, which allows for phenomena like isothermal heat transfer and pumping heat against a temperature gradient. By shifting the paradigm from passive analysis to active control and design, the NEGF formalism opens a new frontier for manipulating thermal energy at the nanoscale, with profound implications for thermal management, energy conversion, and thermal information processing.
\end{justify}
\end{abstract}

\section{Introduction}
Thermal radiation is a fundamental physical process by which heat energy is emitted and exchanged via electromagnetic waves~\cite{add1,add2,add3}. Our understanding of radiative heat transfer (RHT) has long been anchored by Planck's law of blackbody radiation, which sets an upper limit on far-field thermal emission between bodies. In the far-field regime (separations much larger than the thermal wavelength), radiative exchange is well described by blackbody theory~\cite{add4}. However, when objects are brought into close proximity (sub-wavelength gaps), ``near-field radiative heat transfer" can vastly exceed the blackbody limit due to tunneling of evanescent waves~\cite{add5}. The discovery and confirmation of this near-field enhancement --- orders of magnitude larger heat flux than blackbody predictions --- has opened exciting possibilities for nanoscale thermal management and energy conversion~\cite{add6}. Over the past two decades, theoretical and experimental advances have established near-field RHT as a vibrant research field~\cite{add7}, encompassing investigations of surface electromagnetic modes~\cite{add8}, phenomena such as noncontact friction~\cite{add9}, and extensions to complex many-body systems~\cite{add10}. Several comprehensive reviews have chronicled these rapid developments, establishing a solid foundation for the field's principles and applications~\cite{add11,add12}.

Conventional analyses of RHT rely on fluctuational electrodynamics (FE), a framework pioneered by Rytov and others in the mid-20th century~\cite{add13,add14,add15}. FE combines Maxwell's equations with the fluctuation-dissipation theorem (FDT)~\cite{add16} to relate the random thermal currents inside materials to electromagnetic field fluctuations in and around bodies at finite temperature. Using FE, Polder and Van Hove in 1971 presented the first quantitative theory of radiative heat transfer between closely spaced bodies~\cite{add17}, showing that evanescent (non-radiative) field coupling can vastly increase heat flux at nanoscale gaps. Since then, FE-based methods have been extensively developed --- including derivations of the Lifshitz formula for parallel planes and generalizations via scattering matrix formalisms --- to compute near-field RHT for diverse geometries and materials~\cite{add18,add19,add20}. This theoretical framework has not only predicted remarkable near-field enhancements but has also guided numerous experiments validating FE theory. These experiments span various configurations and materials, from pioneering scanning thermal microscope measurements~\cite{add5} to studies between parallel plates~\cite{add21} and in extreme near-field gaps~\cite{add22}, confirming theoretical predictions for surface phonon polariton-mediated transfer~\cite{add23}.

While fluctuational electrodynamics is exceptionally successful for equilibrium situations, it inherently assumes each object is internally in local thermal equilibrium (with a well-defined temperature) and that the radiation exchange is a linear response process governed by equilibrium fluctuation statistics. These assumptions become inadequate in emerging scenarios where systems depart from local equilibrium. Notable examples include materials with time-varying properties found in Floquet systems~\cite{add24} and temporal metamaterials~\cite{add25}, as well as systems under strong external fields requiring advanced non-equilibrium many-body theories~\cite{add26}, which challenge conventional equilibrium thermodynamics~\cite{add27}. Time-modulated~\cite{add28,add29} and current-driven materials~\cite{add30,add31,add32} represent new frontiers beyond FE's scope. Additional challenges emerge in nanogap devices where conduction and radiation coexist and interact~\cite{add33}. Consequently, researchers have increasingly adopted the non-equilibrium Green's function (NEGF) formalism to investigate radiative heat transfer beyond conventional limits~\cite{add34,add35,add36}. The NEGF approach, originating from Keldysh's foundational work~\cite{add37} and comprehensively developed in quantum kinetics textbooks~\cite{add38}, was initially applied to electronic and phononic transport~\cite{add39,add40,add41}. A seminal review by Wang \textit{et al}.~\cite{add34} provides a rigorous theoretical development of NEGF machinery for electron-photon systems, tracing the progression from the quasistatic Coulomb limit to a unified theory in the temporal gauge.
		
We organize our discussion around recent breakthroughs facilitated by NEGF, including the resolution of fluctuational electrodynamics limitations, the unification of radiative and conductive heat channels, the quantum design of thermal metamaterials, and the description of active heat control in driven non-equilibrium systems. Our objective is to establish a conceptual bridge that makes this quantum many-body technique accessible and directly applicable to researchers addressing challenges in nanoscale thermal science. The review proceeds as follows: Section 2 establishes the theoretical foundations of the NEGF formalism for RHT. Section 3 examines recent breakthroughs enabled by this approach. Section 4 explores future research directions and potential applications. Section 5 presents concluding remarks and perspectives on the field's trajectory.

\section{Theoretical foundations}
\subsection{Conventional Approach: Fluctuational Electrodynamics}
To contextualize the non-equilibrium Green's function (NEGF) formalism, we first briefly review the conventional approach to radiative heat transfer: fluctuational electrodynamics (FE).  Fluctuational electrodynamics rests on the idea that thermal radiation originates from thermally excited charge and current fluctuations inside materials~\cite{add14}. In an object at temperature $T$, the random microscopic currents $\mathbf{J}(\mathbf{r},t)$ induce electromagnetic field fluctuations. According to the FDT, the correlation function of these currents in equilibrium is tied to the local temperature and linear response properties (e.g. the imaginary part of the dielectric function) of the material. Rytov's theory formalized this, enabling one to compute statistical properties of the electromagnetic field (e.g. the electric field correlation $\langle E_i E_j^* \rangle$) in terms of dyadic Green's functions of Maxwell's equations and the thermal excitation of sources~\cite{add13}. In essence, the dyadic Green's function $\mathbf{G}(\mathbf{r},\mathbf{r'};\omega)$ relates a point source at $\mathbf{r'}$ to the field response at $\mathbf{r}$; its imaginary part encodes the density of electromagnetic states available for radiative transfer. By integrating these fluctuations via Poynting's theorem, one obtains the radiative heat flux between bodies.

For a canonical system of two bodies (1 and 2) maintained at temperatures $T_1$ and $T_2$ and separated by a vacuum gap, the FE formalism yields a Landauer-like expression for the net radiative heat current, $I$:
\begin{equation}
	I = \int_0^\infty\frac{d\omega}{2\pi}\hbar\omega\left[N_1(\omega)-N_2(\omega)\right]\mathcal{T}(\omega),
\end{equation}
where $N_\alpha(\omega) = 1/(\exp(\hbar\omega/k_B T_\alpha) - 1)$ is the Bose-Einstein distribution at temperature $T_\alpha$, and $\mathcal{T}(\omega)$ is the frequency-dependent transmission (or energy transmission coefficient) for photons across the gap. This Landauer-like form highlights that radiative heat transfer can be viewed as bosonic energy carriers (photons or evanescent modes) transmitting between two thermal reservoirs, weighted by the difference in occupation ($N_1-N_2$) and the available transmission spectrum $\mathcal{T}(\omega)$. In the far-field limit (large separation), $\mathcal{T}(\omega)$ reduces to the overlap of each body's far-field emissivity/absorptivity, recovering Stefan-Boltzmann's law (bounded by $\mathcal{T}\le 1$ per mode, consistent with Planck's limit). In the near-field regime, however, $\mathcal{T}(\omega)$ can greatly exceed the far-field limits for certain frequencies due to contributions of evanescent, tunneling modes. For example, surface polaritons or plasmonic modes can resonantly enhance $\mathcal{T}(\omega)$, yielding net heat flux far above blackbody predictions.

Through either direct solution of Maxwell's equations with fluctuating sources (in simple geometries) or more modern approaches like the scattering matrix formalism, FE can quantitatively predict near-field RHT. Notably, for parallel planar surfaces separated by a gap $d$, Polder and Van Hove's theory showed that as $d$ decreases below the dominant thermal wavelength (typically $10-100~\mu$m at room temperature), the heat flux grows as $\sim 1/d^2$ in the non-retarded limit, due to evanescent electrostatic couplings, until other effects (e.g. material properties, near-contact tunneling) intervene. Many experiments since 2000 have confirmed these predictions, observing near-field heat flux enhancements of 10$^2$--10$^4$ compared to blackbody radiation at sub-micron gaps~\cite{add6,add7}. These validations firmly establish fluctuational electrodynamics as the conventional toolset for analyzing radiative heat exchanges at the nanoscale.

Despite its successes, the FE approach has known limitations. It assumes each object can be assigned a well-defined local temperature and uses equilibrium response functions (dielectric constants, etc.) evaluated at that temperature. This assumption breaks down if an object is internally driven out of equilibrium (for instance, by an external pump or time-varying fields) such that its emission is not simply given by an equilibrium thermal spectrum. Moreover, FE treats radiation in isolation --- it does not inherently include conduction or other channels of heat unless one simply adds them ad hoc. In extreme near-field situations (nanometer or sub-nm gaps), the distinction between radiative and conductive heat transfer can blur, as electrons, phonons, and photons can all interact across the gap. In such regimes, using separate theories for conduction vs. radiation and then summing the results may be inaccurate. Additionally, FE is essentially a linear response theory (relying on FDT), meaning it might not capture novel nonlinear thermal effects or correlated many-body energy exchanges beyond the pairwise interactions of fluctuating dipoles. These challenges motivate alternative approaches that can go beyond the scope of fluctuational electrodynamics.

\subsection{NEGF Formalism for Radiative Heat Transfer}
The non-equilibrium Green's function (NEGF) formalism provides a framework to tackle transport in quantum systems far from equilibrium, and it has been widely employed in electronic and phononic heat conduction studies~\cite{add38}. In essence, NEGF is built on the Keldysh contour technique~\cite{add34}, introducing Green's functions (propagators) for particles (electrons, phonons, etc.) that carry energy between different reservoirs. By including Keldysh lesser Green's functions, NEGF explicitly incorporates the occupation (statistical distribution) of carriers out of equilibrium, rather than assuming they follow equilibrium (FDT) relations. When applying NEGF to radiative heat transfer, one treats thermal photons (or more precisely, electromagnetic field quanta) as the carriers of energy. However, photons are not elementary constituents of matter, so the problem is slightly indirect: radiative transfer arises from the coupling of two material systems via the electromagnetic field.
\subsubsection{Hamiltonian and Choice of Gauge}
A foundational step in constructing a quantum theory of electromagnetism is the choice of gauge. A common choice, the Coulomb gauge ($\nabla \cdot \mathbf{A} = 0$)~\cite{add42}, is physically intuitive as it separates the scalar potential $\phi$, which mediates the instantaneous Coulomb interaction, from the transverse vector potential $\mathbf{A}$, which describes radiation. However, this separation can be a disadvantage for a unified theory of near- and far-field heat transfer, as it necessitates tracking both charge-charge and current-current correlations~\cite{add34}.

In this review, we adopt the temporal gauge, also known as the axial gauge, defined by the condition that the scalar potential is zero ($\phi=0$)~\cite{add43,add44}. This choice is not merely a matter of convenience; it is a strategic decision that leads to a more streamlined and unified theory. By eliminating the scalar potential from the outset, the vector potential $\mathbf{A}$ becomes the sole mediator of the entire electromagnetic interaction. It elegantly encompasses both the longitudinal and transverse components of the field, thereby treating near-field and far-field interactions on an equal footing. This approach is highly economical, as the material's response can be fully characterized by a single quantity: the current-current correlation function, which defines the photon self-energy $\Pi$. Furthermore, the temporal gauge provides a direct and transparent link to the physical, gauge-invariant electric and magnetic fields, which are given by $\bf{E} = \partial\bf{A}/\partial t$ and $\bf{B} = \nabla \times \bf{A}$. This directness is particularly advantageous when comparing the NEGF results to those of classical or fluctuational electrodynamics (FE).

The total Hamiltonian for the coupled system of material electrons and the electromagnetic field can be decomposed into three components:
\begin{equation}
	\hat{H}_{tot} = \hat{H}_{mat} + \hat{H}_{em} + \hat{H}_{int}.
\end{equation}
The material Hamiltonian, $\hat{H}_{mat}$, describes the electronic degrees of freedom. For a system composed of multiple objects ($\alpha = 1, 2, \dots$), we use a tight-binding model where the Hamiltonian for object $\alpha$ is:
\begin{equation}
	\hat{H}_\alpha = \sum_{i,j\in \alpha}c^\dagger_i H_{ij} c_j,
\end{equation}
where $c^\dagger_i$ ($c_j$) creates (annihilates) an electron at site $i$ ($j$) within object $\alpha$.

The Hamiltonian for the free electromagnetic field, $\hat{H}_{em}$, in the temporal gauge is given by~\cite{add45}:
\begin{equation}
	\hat{H}_{em} = \int d\mathbf{r} \left[\frac{\epsilon_0}{2}\left(-\frac{\partial \mathbf{A}}{\partial t}\right)^2 + \frac{1}{2\mu_0}(\nabla\times\mathbf{A})^2\right].
\end{equation}
Here, $\epsilon_0$ and $\mu_0$ are the vacuum permittivity and permeability, respectively. The term $-\partial\mathbf{A}/\partial t$ corresponds to the electric field $\mathbf{E}$ in this gauge.

The interaction between electrons and the field, $\hat{H}_{int}$, is introduced through the Peierls substitution~\cite{add46,add47}. This gauge-invariant procedure modifies the electronic hopping matrix elements to account for the phase acquired by an electron moving in the presence of a vector potential. The combined material and interaction Hamiltonian is then expressed as:
\begin{equation}
	\hat{H}_{mat} + \hat{H}_{int} = \sum_{\alpha} \sum_{i,j\in \alpha}c^\dagger_i H_{ij} c_j \exp\left(\frac{e_0}{i\hbar}\int_{\mathbf{r}_j}^{\mathbf{r}_i}\mathbf{A}\cdot d\mathbf{r}\right),
\end{equation}
where $e_0$ is the elementary charge. The total Hamiltonian is the sum of this term and the electromagnetic field Hamiltonian $\hat{H}_{em}$.

\subsubsection{Photonic Green's Functions and Self-Energies}

The power of the NEGF formalism lies in its systematic treatment of interactions through Green's functions (propagators) and self-energies (interaction terms). For the coupled matter-photon system, we define these quantities for the electromagnetic field.

\paragraph{Photon Green's Function ($\mathbf{D}$)}
The photon Green's function is the central object that describes the propagation of electromagnetic excitations. In the NEGF formalism, this is a matrix of contour-ordered correlation functions of the vector potential operators:
\begin{equation}
	D_{\mu\nu}(\mathbf{r}\tau; \mathbf{r}'\tau') = -\frac{i}{\hbar} \langle T_c \, A_\mu(\mathbf{r}, \tau) A_\nu(\mathbf{r}', \tau') \rangle.
\end{equation}
Here, $\tau$ and $\tau'$ are time variables on the Keldysh contour, $T_c$ is the contour-ordering operator, and the indices $\mu, \nu$ denote Cartesian components. This is the ``dressed" or full Green's function, as it implicitly includes all interaction effects between the electromagnetic field and the material system.

\paragraph{Photon Self-Energy ($\mathbf{\Pi}$)}
The photon self-energy, $\mathbf{\Pi}$, encapsulates the influence of the material system on the photons. It represents the material's response to the electromagnetic field and acts as the source of all absorption, emission, and scattering processes. Formally, it is defined via the irreducible current-current correlation function~\cite{add48,add49}:
\begin{equation}
	\Pi_{\mu\nu}(\mathbf{r}\tau; \mathbf{r}'\tau') = -\frac{i}{\hbar} \langle T_c \, j_\mu(\mathbf{r}, \tau) j_\nu(\mathbf{r}', \tau') \rangle_{\text{ir}}.
\end{equation}
The subscript ``ir" denotes that only irreducible diagrams are included-those that cannot be split into two by cutting a single bare photon line. In a discrete basis (e.g., the tight-binding sites), $\mathbf{\Pi}$ becomes a matrix whose elements $\Pi_{ij}$ describe the current-current correlation between sites $i$ and $j$.

These two quantities are linked by a set of fundamental equations. On the Keldysh contour, the Dyson equation relates the dressed Green's function $\mathbf{D}$ to the free-photon Green's function $\mathbf{v}$ and the self-energy $\mathbf{\Pi}$ in a compact operator notation~\cite{add50}:
\begin{equation}
	\label{dyson}
	\mathbf{D} = \mathbf{v} + \mathbf{v} \mathbf{\Pi} \mathbf{D}.
\end{equation}
Using the Langreth rules~\cite{add51}, this single contour equation can be projected into a set of real-time equations for the retarded, advanced, lesser, and greater components.

\subsubsection{The Free-Photon Green's Function}

To solve the Dyson equation for the full photon propagator $\mathbf{D}$, we require an explicit expression for the free-photon Green's function, $\mathbf{v}$. This function, also known as the bare propagator, describes the propagation of photons in vacuum and is the fundamental solution to the free-field wave equation. For the electromagnetic field, the commutation relation between the vector potential and its conjugate momentum is~\cite{add34}:
\begin{equation}
	[A_\mu(\mathbf{r}), \epsilon_0\dot{A}_\nu(\mathbf{r}')] = i\hbar \delta_{\mu\nu} \delta(\mathbf{r}-\mathbf{r}'),
\end{equation}
where $\dot{A}_\nu(\mathbf{r}')$ is directly proportional to the electric field operator. The use of the standard Dirac $\delta$ function, rather than the transverse $\delta$ function, reflects the fact that $\mathbf{A}$ is a general three-component vector field. The electronic operators obey the standard fermionic anti-commutation relations.

With the Hamiltonian and commutation relations defined, the dynamics of the system are given by the Heisenberg equations of motion. For the vector potential, this yields the quantum wave equation:
\begin{equation}
	\left( \frac{1}{c^2}\frac{\partial^2}{\partial t^2} - \nabla \times (\nabla \times) \right) \mathbf{A}(\mathbf{r}, t) = \mu_0 \mathbf{j}(\mathbf{r}, t),
	\label{Heisenberg}
\end{equation}
with the current operator given by
\begin{equation}
	\mathbf{j}(\mathbf{r}, t) = \frac{1}{i\hbar}\left[ \epsilon_0\dot{\mathbf{A}}(\mathbf{r}),\sum_{i,j}c^\dagger_iH_{ij}c_j\exp\left(\frac{e_0}{i\hbar}\int_{\mathbf{r}_j}^{\mathbf{r}_i}\bf{A}\cdot d\bf{r}\right)\right],
\end{equation}
where the source term, $\mathbf{j}(\mathbf{r}, t)$, is the fully quantum-mechanical current density operator, derived from the variation of the Hamiltonian with respect to $\mathbf{A}$. For the electrons, the Heisenberg equation yields a Schr\"{o}dinger-like equation where the dynamics are governed by the field-dependent Peierls Hamiltonian. This coupled set of equations provides a complete and self-consistent description of the interacting electron-photon system, laying the groundwork for the application of the NEGF method.

To solve the Dyson equation, we need an explicit expression for the free-photon Green's function $v$, especially the retard bare propagator $v^r$. It is the Green's function of the free-field wave equation represents the propagation of photons in a vacuum. The most convenient way to obtain $v$ is by solving Eq.~($\ref{Heisenberg}$) in the frequency ($\omega$) and momentum ($\mathbf{q}$) domain, where the differential operators become algebraic terms. The retarded free Green's function in Fourier space is given by:
\begin{equation}
	\mathbf{v}^r(\mathbf{q}, \omega) = -\frac{1}{\epsilon_0} \frac{\mathbf{U} - \mathbf{q}\mathbf{q}/(k_0^2)}{(\omega+i\eta)^2 - c^2q^2},
\end{equation}
where $k_0=\omega/c$ is the free-space wave number, $\mathbf{U}$ is the 3$\times$3 identity dyadic, $\mathbf{q}\mathbf{q}$ is the dyadic product of the momentum vector, and $\eta$ is a positive infinitesimal that ensures causality. By performing an inverse Fourier transform, we obtain the well-known expression for the dyadic Green's function in real space~\cite{add52,add53}:
\begin{equation}
	\mathbf{v}^r(\mathbf{r}, \omega) = -\frac{\mu_0e^{ik_0 r}}{4\pi r} \left[ \left( \mathbf{U} - \hat{\mathbf{r}}\hat{\mathbf{r}} \right) + \frac{1}{ik_0 r} \left(1 - \frac{1}{ik_0 r}\right) (\mathbf{U} - 3\hat{\mathbf{r}}\hat{\mathbf{r}}) \right],
\end{equation}
where $r = |\mathbf{r}|$ and $\hat{\mathbf{r}} = \mathbf{r}/r$. This expression for $v^r$ serves as the kernel in the Dyson equation, allowing for the calculation of the full Green's function $D^r$ once the material properties contained in $\Pi^r$ are known.

\subsubsection{Calculating the Photon Self-Energy}

With the free propagator $\mathbf{v}^r$ defined, the remaining task is to determine the photon self-energy $\mathbf{\Pi}$, which contains all the material-specific information. The self-energy is derived from the electron-photon interaction term in the Hamiltonian and can be calculated using diagrammatic perturbation theory~\cite{add54,add55}.

A common and powerful approach for systems where electron-electron interactions are important is the Random Phase Approximation (RPA)~\cite{add56}. Within RPA, the photon self-energy is given by the lowest-order irreducible polarization bubble diagram. For a discrete tight-binding model, this translates to~\cite{add32}:
\begin{equation}
	\Pi^{RPA}_{l\mu,l'\nu}(\tau,\tau')=-i\hbar\sum_{ijkp}M^{l\mu}_{ij}G_{jk}(\tau,\tau')M^{l'\mu}_{kp}G_{pi}(\tau',\tau),
\end{equation}
where $M_{ij}^{l\mu} = e(\delta_{il}+\delta_{jl})v^\mu_{ij}/2$ is the local current operator between site $i$ and $j$ with $v^\mu$ is the component of the velocity matrix in the $\mu$ direction. This expression has a clear physical meaning: the self-energy describes the process where a photon is absorbed by creating an electron-hole pair (represented by the product of two electron Green's functions), which then propagates and subsequently recombines. This process is what gives rise to the material's polarization and dissipation.

\subsubsection{Derivation of the Heat Current: The Meir-Wingreen Formula}

Having established the NEGF machinery, we are now equipped to derive an expression for the radiative heat current. The derivation provides a powerful link between the macroscopic law of energy conservation and the microscopic quantum Green's functions.

The starting point is the law of energy conservation for an electromagnetic field interacting with matter, given by Poynting's theorem~\cite{add57}:
\begin{equation}
	\frac{\partial u_{em}}{\partial t} + \nabla \cdot \mathbf{S} = -\mathbf{E} \cdot \mathbf{j},
\end{equation}
where $u_{em}$ is the energy density of the field, $\mathbf{S}$ is the Poynting vector (energy flux), and the term on the right-hand side, $-\mathbf{E} \cdot \mathbf{j}$, is the power per unit volume transferred from the field to the material currents.

In a non-equilibrium steady state, the time derivative of any ensemble-averaged quantity is zero. Integrating over the volume $V_\alpha$ of a specific object $\alpha$ and applying the divergence theorem, we find that the net power emitted by the object, $I_\alpha$, is equal to the total power dissipated by the currents within that volume:
\begin{equation}
	I_\alpha = \oint_{\Sigma_\alpha} \langle \mathbf{S} \rangle \cdot d\mathbf{\Sigma} = \int_{V_\alpha} \langle -\mathbf{E} \cdot \mathbf{j}_\alpha \rangle dV.
\end{equation}
Here, $\mathbf{j}_\alpha$ is the current operator associated only with object $\alpha$. The central task is to express the integrand, the power density, using the NEGF quantities. In the temporal gauge ($\mathbf{E} = -\partial_t \mathbf{A}$), the power emitted by object $\alpha$ becomes:
\begin{equation}
	I_\alpha = \int_{V_\alpha} \langle (\partial_t \mathbf{A}) \cdot \mathbf{j}_\alpha \rangle dV.
\end{equation}
This expression reveals that all transport quantities are expectation values of products of the field operator $\mathbf{A}$ (or its derivatives) and the current operator $\mathbf{j}$. This structure motivates the definition of an intermediate Green's function, $F$, which correlates the field with the material current~\cite{add34}:
\begin{equation}
	F_{\mu\nu}(\mathbf{r}\tau; \mathbf{r}'\tau') = -\frac{i}{\hbar} \langle T_\tau A_\mu(\mathbf{r}, \tau) j_\nu(\mathbf{r}', \tau') \rangle.
\end{equation}
Using this definition and taking the symmetrically-ordered quantum expectation value, we can express the total transported quantities as:
\begin{equation}
	I_\alpha = \rm{Re} \int_0^\infty \frac{d\omega}{2\pi} \hbar\omega \rm{Tr} \left[ \mathbf{F}^K_\alpha(\omega) \right],
\end{equation}
where Tr[...] includes an integral over the volume of object $\alpha$ and a trace over the dyadic indices.

The final, crucial step is to relate $F$ back to the fundamental NEGF quantities $D$ and $\Pi$. A rigorous proof based on the Dyson equation and the operator equations of motion reveals a remarkably simple and powerful identity~\cite{add32,add34}:
\begin{equation}
	\mathbf{F}_\alpha = -\mathbf{D} \mathbf{\Pi}_\alpha.
\end{equation}
This relation is exact, provided one accepts the additivity of the self-energy, $\Pi = \sum_{\alpha}\Pi_{\alpha}$, which is an excellent approximation at the level of the RPA. This identity is the linchpin of the theory: it states that the response of object $\alpha$ generates a field that propagates throughout the entire system $D$.

Substituting this identity into our expressions and using the Langreth rule for the Keldysh component, $F^K_\alpha = -(D^r\Pi^K_\alpha+D^K\Pi^a_\alpha)$, we arrive at the unified Meir-Wingreen formulas for transport~\cite{add58,add59}:
\begin{equation}
	\label{M-W}
	I_\alpha = \int_0^\infty \frac{d\omega}{2\pi} (-\hbar\omega) \rm{Re Tr} \left[ \mathbf{D}^r \mathbf{\Pi}^K_\alpha + \mathbf{D}^K \mathbf{\Pi}^a_\alpha \right].
\end{equation}
Because of the real part in Eq.~(\ref{M-W}), we have a more symmetric form by adding the Hermitian conjugate inside the trace and divided by 2, using the general relations, $D^r - D^a = D^>-D^<$, $D^K = D^>+D^<$, and similarly for the self-energies $\Pi$, we have:
\begin{equation}
	I_\alpha = \int_0^\infty \frac{d\omega}{2\pi} (\hbar\omega) \rm{Tr} \left[ \mathbf{D}^< \mathbf{\Pi}^>_\alpha-\mathbf{D}^> \mathbf{\Pi}^<_\alpha \right].
\end{equation}
This formula represent the central achievement of the NEGF transport theory. This powerful formula has a clear physical meaning: the net heat current is determined by the balance between energy emission from object $\alpha$ (proportional to its `greater' self-energy $\Pi^>_\alpha$) and energy absorption by it (proportional to its `lesser' self-energy $\Pi^<_\alpha$), where each process is mediated by the full system's photon propagator ($D^<$ and $D^>$).

To illustrate the utility and power of these results, we will now connect them to more familiar theories by considering the local equilibrium limit. Subsequently, we will discuss how the NEGF formalism can be integrated with first-principles electronic structure methods to enable quantitative, parameter-free predictions of radiative heat transfer in complex scenarios.

\subsection{Local Equilibrium: Equivalence with Fluctuational Electrodynamics}

A crucial test for any general non-equilibrium theory is its ability to recover well-established equilibrium results. When each object $\alpha$ in the system can be considered to be in its own state of local thermal equilibrium at a temperature $T_{\alpha}$, the NEGF formalism must reproduce the results of Rytov's fluctuational electrodynamics.

In the language of NEGF, local thermal equilibrium is expressed through the Fluctuation-Dissipation Theorem. The FDT states that for an object in thermal equilibrium, the quantum fluctuations (related to the Keldysh component, $\Pi^K$) are intrinsically linked to the system's dissipative response (related to the retarded component, $\Pi^r$). For the photon self-energy of object $\alpha$, the FDT takes the form:
\begin{equation}
	\mathbf{\Pi}^K_\alpha(\omega) = (2N_\alpha(\omega) + 1) \left[ \mathbf{\Pi}^r_\alpha(\omega) - \mathbf{\Pi}^a_\alpha(\omega) \right].
\end{equation}
The term ($2N_\alpha + 1$) contains both thermal fluctuations ($2N_\alpha$) and quantum vacuum fluctuations (+1).

By substituting the FDT into the Meir-Wingreen formula for energy current ($I_\alpha$) and using the Keldysh equation to express $D^K$, a remarkable simplification occurs. After algebraic manipulation, all terms involving the vacuum fluctuations (+1) perfectly cancel out. The final expression for the net energy current flowing out of object $\alpha$ reduces to the celebrated Landauer-B\"{u}ttiker formula:
\begin{equation}
	I_\alpha = \sum_{\beta} \int_0^\infty \frac{d\omega}{2\pi} \hbar\omega \, \mathcal{T}_{\beta\alpha}(\omega) \left[ N_\beta(\omega) - N_\alpha(\omega) \right],
\end{equation}
where the transmission coefficient, $\mathcal{T}_{\beta\alpha}(\omega)$, is given by the Caroli formula~\cite{add60,add61,add62}:
\begin{equation}
	\label{caroli}
	\mathcal{T}_{\beta\alpha}(\omega) = \rm{Tr} \left[ \mathbf{D}^r(\omega) \mathbf{\Gamma}_\beta(\omega) \mathbf{D}^a(\omega) \mathbf{\Gamma}_\alpha(\omega) \right],
\end{equation}
with spectrum function $\Gamma_\alpha = i(\Pi^r_\alpha-\Pi^a_\alpha)$ representing the dissipation rate into bath $\alpha$. This formula is formally identical to the expressions for heat transfer derived from fluctuational electrodynamics, thus proving the equivalence of the two theories in the local equilibrium limit.

\subsubsection{Example: Recovering the Polder-van Hove Formula}

To make this equivalence concrete, we apply the NEGF result to the canonical problem of heat transfer between two large, parallel plates (1 and 2) separated by a vacuum gap of distance $d$. The goal is to derive the well-known frequency- and wavevector-dependent transmission coefficient.

For this planar geometry, it is convenient to work in a mixed representation where we perform a 2D Fourier transform in the parallel plane $(x,y) \to (q_x, q_y)$ but remain in real space for the perpendicular direction, $z$. Due to the symmetry of the problem, the electromagnetic modes decouple into two independent polarizations: s-polarization (TE) and p-polarization (TM). We can therefore perform the calculation for each polarization separately and sum the results.

For a given in-plane wavevector $\mathbf{q}$ and polarization state, the key quantities in Eq.~(\ref{caroli}) are the dressed Green's functions $D^r$ and $D^a$, and the coupling matrices $\Gamma_1$ and $\Gamma_2$. The self-energy $\Pi_\alpha$ is a local property of material $\alpha$. In the scattering picture, it is directly related to the Fresnel reflection coefficient $r_\alpha(\omega, \mathbf{q})$ of the semi-infinite plate $\alpha$. The coupling matrix $\Gamma_\alpha$ is then related to the material's absorptivity.

Since $D^a = (D^r)^\dagger$ and $\Pi^a = (\Pi^r)^\dagger$, in the following discussion, we omit the superscript, and all quantities are retarded unless otherwise specified. For two parallel plates system, it is convenient to partition Green's function and self-energy as block matrices:
\begin{equation}
D = \left[
\begin{array}{ll}
		D_{11} & D_{12}\\
		D_{21} & D_{22}
\end{array}
\right],
v = \left[
\begin{array}{ll}
	v_{11} & v_{12}\\
	v_{21} & v_{22}
\end{array}
\right],
\Pi = \left[
\begin{array}{ll}
	\Pi_{11} & 0\\
	0 & \Pi_{22}
\end{array}
\right].
\end{equation}
In other words, each diagonal block of the matrix consists of localized quantities of each plate in isolation and all entries in the off-diagonal block are interaction terms between two plates.
Taking the trace of each side explicitly, the transmission coefficient of Eq.~(\ref{caroli}) can be simplified as~\cite{add63}
\begin{equation}
\label{caroli3}
\mathcal{T}(\omega)=\int\frac{d^2{\bf q}}{(2\pi)^2}\big[D_{21}\Gamma_{11} D_{21}^\dagger\Gamma_{22}\big],
\end{equation}
where all quantities are in the 2D form and ${\bf q} = (q_x, q_y)$ lies in the first Brillouin zone.

For planar geometry, the free Green's function can be expressed with the mixed representation, for which we Fourier transform the variable in $z$ direction back to real space, with the result
\begin{equation}
	\mathbf{v}^r(\mathbf{q},z,z' \omega) = \frac{\mu_0e^{iq_z|z-z'|}}{2ik^2_0q_z}\left[
	\begin{array}{ccccc}
		k^2_0-q^2_x&-q_xq_y&-sq_xq_z\\
		-q_xq_y&k^2_0-q^2_y&-sq_yq_z\\
		-sq_xq_z&-sq_yq_z&k^2_0-q^2_z\\
	\end{array}\right],
\end{equation}
where s is the sign of ($z-z'$), and $q_z = \sqrt{k^2_0-q^2}$ for the propagating mode when $k_0 > q$ and $q_z = i\sqrt{q^2-k^2_0}$ for the evanescent mode when $k_0<q$.
Because the $s$ and $p$ channels of the electromagnetic waves are independent, the free photon Green's function can be further expressed by $v = (i/2q_z)[\hat{\mathbf{e}}_s\hat{\mathbf{e}}_s+\hat{\mathbf{e}}_p\hat{\mathbf{e}}_p]\exp(iq_z|z-z'|)$, where $\hat{\mathbf{e}}_s$ and $\hat{\mathbf{e}}_p$ are unit vectors for $s$ and $p$ polarization, respectively. In this case, the final expression is the same for $s$ and $p$ polarization.

Since $|z-z'|= d$, we can introduce $v_q$ such that $v_{11} = v_{22} = v_q$ and $v_{12} = v_{21} = v_qe^{iq_zd}$. While a full first-principles calculation of the self-energy $\Pi$ is complex, its effect in this planar geometry can be conveniently encapsulated. For a semi-infinite body, the retarded surface photon self-energy is directly related to the Fresnel reflection coefficient $r$ by $v_q\Pi_{\alpha\alpha} = r_\alpha/(1+r_\alpha)$~\cite{add63,add64}.

The off-diagonal block of the dressed Green’s function $D^r_{21}$ describes the propagation from plate 2 to plate 1, including all multiple reflections within the Fabry-P\'{e}rot-like cavity formed by the two plates. From Dyson equation (Eq.~(\ref{dyson})), this can be calculated by summing a geometric series of scattering events~\cite{add62,add63}:
\begin{equation}
	D_{21} = \frac{(1+r_1)(1+r_2)v_qe^{iq_zd}}{1-r_1r_2e^{2iq_zd}}.
\end{equation}
With the spectrum function $\Gamma=i(\Pi-\Pi^\dagger)$, the transmission coefficient Eq.~(\ref{caroli3}) finally becomes:
\begin{equation}
	\mathcal{T}(\omega, q) =
	\begin{cases}
		\frac{(1 - |r_1|^2)(1 - |r_2|^2)}{|1 - r_1 r_2 e^{2iq_z d}|^2}, & q < k_0 \\
		\frac{4 \text{Im}(r_1) \text{Im}(r_2) e^{-2|\text{Im}(q_z)| d}}{|1 - r_1 r_2 e^{-2|\text{Im}(q_z)| d}|^2}, & q > k_0
	\end{cases}
\end{equation}
This is precisely the celebrated result first derived by Polder and van Hove using fluctuational electrodynamics. The top case describes the contribution of propagating waves, which are limited by emissivity $(1-|r|^2)$, while the bottom case describes the tunneling of evanescent waves, which dominates in the near-field. This successful recovery of the canonical FE result firmly establishes the validity and generality of the NEGF formalism.

\subsection{Specialization to the Near-Field and First-Principles Inputs}

The true power of the NEGF formalism lies in its generality. In this section, we explore two special frontiers: the specialization of the theory to the quasistatic (near-field) limit, where Coulombic interactions dominate, and the integration with first-principles electronic structure methods to enable parameter-free predictions~\cite{add65}.

\subsubsection{The Quasistatic Limit: A Scalar Field Description}
In above sections, we introduced the general NEGF theoretical framework for RHT in the temporal gauge ($\phi = 0$). With the full vector potential $\mathbf{A}$, the near-field and far-field regime can be treated on an equal footing. However, in the extreme near field ($d \ll \lambda_T$, where $\lambda_T$ is the thermal wavelength), the retardation effects included in the full vector potential formalism become less important. The energy transfer is dominated by the coupling of evanescent electromagnetic modes, which can be viewed as the fluctuations of the scalar potential ($\phi$) associated with charge density fluctuations ($\rho$)~\cite{add66,add67,add68}.

This insight allows for a significant simplification of the theory by taking the quasistatic limit ($c \to \infty$). In this limit, which corresponds to adopting the Coulomb gauge and neglecting the vector potential, the interaction is mediated by the instantaneous Coulomb force~\cite{add69,add70}. The relevant field is the scalar potential $\phi$, and its source is the charge density $\rho$. The NEGF formalism for RHT remains structurally identical, but the key quantities are re-interpreted:

1. Field and Source: The fundamental field is $\phi(\mathbf{r}, \tau)$ and the source is the charge density operator $\rho(\mathbf{r}, \tau)$.

2. Green's Function and Self-Energy: The ``photon" Green's function now describes the propagation of scalar potential fluctuations, $D(\mathbf{r}\tau; \mathbf{r}'\tau') = -i\langle T_c \phi(\mathbf{r}, \tau) \phi(\mathbf{r}', \tau') \rangle$. The self-energy, $\Pi$, becomes the irreducible charge-charge correlation function, $\Pi(\mathbf{r}\tau; \mathbf{r}'\tau') = -i\langle T_c \rho(\mathbf{r}, \tau) \rho(\mathbf{r}', \tau') \rangle_{\text{ir}}$. This quantity is precisely the electronic polarizability tensor.

3. Free Propagator: The free propagator, $v$, is no longer the full dyadic Green's function but simply the bare Coulomb interaction: $v(\mathbf{r}, \mathbf{r}') = 1/(4\pi\epsilon_0|\mathbf{r}-\mathbf{r}'|)$, which in Fourier space becomes $v(\mathbf{q}) = 1/(\epsilon_0 q^2)$ in 3D and $1/(2\epsilon_0 q$) in 2D systems.

Remarkably, with these re-interpretations, the core NEGF equations-the Dyson equation ($\mathbf{D}^r = \mathbf{v}^r+\mathbf{v}^r \mathbf{\Pi}^r\mathbf{D}$), the Keldysh equation, and the Meir-Wingreen formulas-all retain their exact form. This demonstrates the robustness of the NEGF framework; it provides a unified language for describing both fully retarded radiative transfer and quasistatic Coulombic transfer~\cite{add34}.

\subsubsection{Self-Energies from First Principles}
To move from a formal theory to quantitative prediction, we need a method to calculate the material-specific photon self-energy, $\mathbf{\Pi}$. Besides the tight-binding method as introduced above, this essential input can also be obtained from the first-principles method developed in modern computational materials science.

The electron Green's function on the Keldysh contour is defined as:
\begin{equation}
	G_{ij}(\tau,\tau') = -i\hbar\langle T_\tau c_i(\tau)c^\dagger_j(\tau') \rangle.
\end{equation}
Under the RPA, the self-energy is the product of the electron Green's function:
\begin{equation}
	\Pi(\mathbf{r}\tau; \mathbf{r}'\tau') =-i\hbar G_{ij}(\tau,\tau')G_{ji}(\tau',\tau).
\end{equation}
Using first-principles methods, we can compute the electronic structure of a material and its response to electromagnetic fields. The most common approach starts with Density Functional Theory (DFT)~\cite{add71,add72}. The central quantity to compute is the irreducible polarizability, $\Pi^0$, which describes the response of non-interacting Kohn-Sham electrons to a change in the potential. Within the RPA, $\Pi^0$ is given by the Adler-Wiser formula~\cite{add73,add74}:
\begin{equation}
	\label{eq:adler-wiser}
	\Pi^0_{\mathbf{G}\mathbf{G}'}(\mathbf{q},\omega) = \frac{2}{V} \sum_{n,n',\mathbf{k}} \frac{f_{n\mathbf{k}} - f_{n'\mathbf{k+q}}}{\epsilon_{n\mathbf{k}} - \epsilon_{n'\mathbf{k+q}} + \hbar\omega + i\eta} \times M_{nn'}(\mathbf{k},\mathbf{q},\mathbf{G}) M^\dagger_{n'n}(\mathbf{k},\mathbf{q},\mathbf{G}'),
\end{equation}
where $V$ is the volume of the system and $M_{nn'}(\mathbf{k},\mathbf{q},\mathbf{G}) = \langle u_{n\mathbf{k}} | e^{-i(\mathbf{q}+\mathbf{G})\cdot\mathbf{r}} | u_{n'\mathbf{k+q}} \rangle$ is the matrix element between Kohn-Sham cell-periodic wavefunctions $u_{n\mathbf{k}}$, $\epsilon_{n\mathbf{k}}$ are the Kohn-Sham eigenvalues, $f_{n\mathbf{k}}$ is the Fermi-Dirac occupation, and $\mathbf{G}, \mathbf{G}'$ are reciprocal lattice vectors that account for microscopic field variations.

This DFT+RPA approach provides a powerful, parameter-free method for predicting radiative heat transfer~\cite{add63,add71}. Furthermore, it can be systematically improved. The well-known band-gap problem of DFT can be corrected by replacing the Kohn-Sham energies $\epsilon_{n\mathbf{k}}$ in Eq.~(\ref{eq:adler-wiser}) with more accurate quasiparticle energies obtained from the GW method (GW+RPA)~\cite{add75}. Moreover, the RPA neglects excitonic effects~\cite{add76} (the attractive interaction between the excited electron and the hole it leaves behind). These can be included by solving the Bethe-Salpeter equation~\cite{add77}, which builds upon the GW results to provide a highly accurate description of the optical response and, therefore, the photon self-energy (GW+BSE). This synergy between the NEGF transport formalism and first-principles calculations opens the door to the predictive modeling of near-field heat transfer in real materials and complex nanostructures.

\section{Recent Advances Enabled by NEGF in RHT}
\subsection{Beyond Local Approximations: A Quantum-Accurate View of Equilibrium RHT}

A primary motivation for applying the Non-Equilibrium Green's Function (NEGF) formalism to radiative heat transfer (RHT) was to transcend the limitations of local approximations inherent in conventional fluctuational electrodynamics (FE). While immensely successful, FE typically relies on macroscopic, local dielectric functions ($\epsilon$) to describe a material's response. This approximation breaks down at the nanoscale, where the response to an electric field at a point $\mathbf{r}$ depends on the field in a surrounding region, a phenomenon known as non-locality. Furthermore, local models can lead to unphysical predictions, such as a divergent heat flux between two surfaces as their separation $d$ approaches zero. The NEGF formalism, by building the material's response from its fundamental quantum-mechanical constituents, naturally incorporates these crucial non-local and quantum effects, enabling a more accurate and physically sound description of RHT even in local thermal equilibrium.

Graphene, with its unique two-dimensional electronic structure, has served as the canonical system for demonstrating these capabilities~\cite{add64,add78}. Early NEGF studies by Jiang and Wang~\cite{add61} based on tight-binding Hamiltonians provided a unified description of near-field RHT (NFRHT) between two graphene sheets, seamlessly spanning from atomic contact to micron-scale separations. Crucially, this approach resolved the problem of the $1/d$ divergence predicted by local theories~\cite{add79,add80}. The NEGF calculations showed that the heat flux naturally saturates at a finite value in the contact limit ($d\to 0$), reaching an enhancement of $\sim 5\times 10^4$ times the blackbody limit (Fig.~\ref{fig1}a). This saturation is a direct consequence of the finite wave-vector components of the electronic response, an effect absent in local models \cite{add63}. It is important to clarify the physical meaning of this saturated flux. It represents the maximum possible energy transfer rate between two distinct bodies mediated by near-field Coulomb interactions across a vacuum gap. This value should not be confused with the bulk thermal conductivity of the material itself. The latter describes heat transport within a continuous medium via propagating carriers (e.g., electrons or phonons), whereas the former describes energy tunneling across a vacuum interface via evanescent electromagnetic fields.

Furthermore, these studies revealed a nuanced power-law decay around $I \propto d^{-2}$ at larger separations, consistent with the power-law expected for an ideal 2D conductor~\cite{add81,add82}. The subsequent development of a parameter-free NEGF framework by Zhu \textit{et al}~\cite{add63,add71}, integrated with first-principles DFT, solidified these findings. By computing the polarizability directly from the Adler-Wiser formula using Kohn-Sham wavefunctions, this method eliminated empirical parameters and quantitatively reproduced the key behaviors (see Fig.~\ref{fig1}b), establishing NEGF as a truly predictive tool for RHT in low-dimensional materials. Figure~\ref{fig1}b also shown the equivalence between FE and NEGF in local thermal equilibrium.

\begin{figure}

	\centering
	\includegraphics[width=1.0\textwidth]{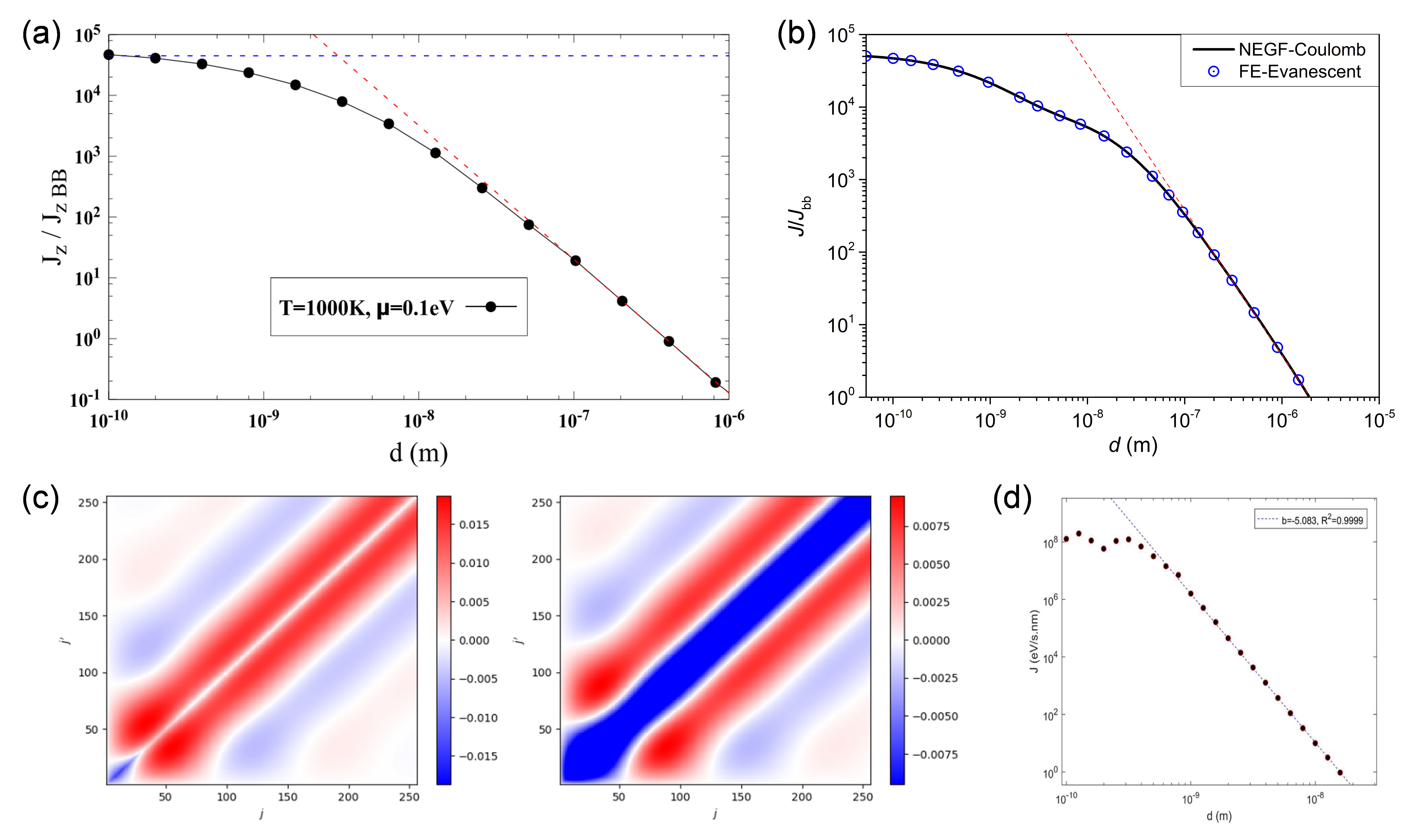}
	\captionsetup{justification=justified, singlelinecheck=false}\caption{(a) Heat flux between two parallel graphene sheets as a function of distance. The NEGF calculation (solid line) shows a saturation of heat flux at short distances, resolving the unphysical $1/d$ divergence predicted by local theories. At larger distances, the flux follows a $\sim d^{-2}$ scaling law (red dashed line). (b) The NEGF and FE equivalence. This conceptual diagram illustrates that NFRHT, described as the tunneling of evanescent waves in fluctuational electrodynamics, is equivalently described in the NEGF formalism as energy transfer mediated by the Coulomb interactions between quantum charge fluctuations in the two bodies. The response function is obtained from first-principles DFT calculations. (c) Visualization of non-local response at a metal surface. The plot shows the real part of the normal component of the polarizability, $\rm{Re}[\Pi_{zz}(j,j')]$, as a function of lattice layer indices $j$ and $j'$. The response is highly non-diagonal, extending over $\sim 100$ atomic layers from the surface ($j = 1$), demonstrating the failure of local models which would be entirely diagonal. (d) Dimensional dependence of heat transfer. The heat flux in a line-surface (1D-2D) configuration decays with distance as $\sim d^{-5}$, a significantly steeper dependence than the $\sim d^{-2}$ law for a surface-surface (2D-2D) system. This highlights how NEGF captures the profound influence of electronic dimensionality on RHT. Panel (a) reprinted with permission from~\cite{add61}. Copyright 2017 American Physical Society. Panel (c) reprinted with permission from~\cite{add83}. Copyright 2025 American Physical Society. Panel (d) reprinted with permission from~\cite{add84}. Copyright 2020 Springer Nature.}
	\label{fig1}

\end{figure}

The power of NEGF to resolve quantum-scale non-locality extends beyond 2D materials to the surfaces of bulk metals, where the classical Drude model fails~\cite{add83}. Rigorous NEGF calculations using surface Green's function techniques have dissected the charge and current response functions of a semi-infinite metal block layer by layer. These studies revealed that while the transverse response ($\Pi_{xx/yy}$) approximates the Drude model, the normal component ($\Pi_{zz}$) exhibits profound non-local characteristics. As shown in Fig.~\ref{fig1}c, the correlation between layers $j$ and $j'$ extends deep into the bulk, over a range of $|j - j'| \sim 100$ atomic layers, with a slow, oscillatory decay. This demonstrates that the electronic response to a field at the surface is a collective, non-local phenomenon involving electrons many layers deep, a reality completely missed by any local or surface-response model.

This sensitivity to the underlying quantum structure means that NEGF can also transparently capture how RHT is governed by the system's geometric dimensionality. When the configuration is changed from two parallel 2D surfaces to a 1D nanotube interacting with a 2D surface, the spatial decay of heat flux is dramatically enhanced. NEGF calculations show that the scaling law steepens from $I \propto d^{-2}$ for the 2D-2D case to $I \propto d^{-5}$ for the 1D-2D configuration (Fig.~\ref{fig1}d)~\cite{add84}. This striking change originates from the fundamentally different spatial correlation of charge fluctuations imposed by the confined electronic states of the 1D system. 

In essence, the NEGF formalism directly links the dimensionality of the quantum states to the macroscopic heat transfer behavior, providing insights inaccessible to models that do not resolve the system's electronic structure. Together, these examples illustrate that by treating the material response from first principles, the NEGF method provides a quantum-accurate view of equilibrium RHT, correcting unphysical divergences and revealing a rich landscape of non-local and dimensional effects that are paramount at the nanoscale.

\subsection{Unifying Heat Transfer Channels: The Synergy of Photons, Electrons, and Phonons}

At the nanoscale, and particularly in sub-nanometer gaps, the classical distinction between heat transfer mechanisms---radiation, conduction, and convection---becomes fundamentally blurred. Energy can be carried simultaneously by photons, electrons, and phonons, and these transport channels can interact and compete in highly non-trivial ways. A signal achievement of the NEGF formalism is its ability to provide a unified quantum-mechanical framework that treats these different energy carriers on an equal footing. By starting from a total Hamiltonian that includes all relevant particles and their interactions, NEGF can naturally describe their synergistic behavior, revealing a complex energy transfer landscape that cannot be captured by simply summing the contributions from separate, independent theories.

The synergy between radiative and electronic transport is starkly illustrated in metal-vacuum-metal (M-V-M) junctions at sub-nanometer separations. In this regime, energy can be transferred via near-field Coulomb fluctuations (the microscopic origin of NFRHT) and by the direct quantum tunneling of electrons across the vacuum barrier. Using a unified NEGF model, Zhang \textit{et al}.~\cite{add62} quantified the crossover between these two mechanisms. As shown in Fig.~\ref{fig2}a, for gap distances $d>0.92$ nm, the heat flux is dominated by Coulomb fluctuations and follows the familiar $d^{-2}$ scaling law of NFRHT. However, as the gap narrows further, electron tunneling effects increase exponentially and become the dominant heat transfer channel, leading to a total flux that is orders of magnitude greater than the radiative contribution alone. Crucially, the study revealed that the two channels are not merely additive. In the strong tunneling regime ($d < 0.6$ nm), the presence of electron tunneling actually enhances the Coulomb-mediated heat flux~\cite{add85}, indicating a non-linear coupling between the two quantum processes. This finding, which helps explain experimentally observed ``abnormally high" thermal conductances, underscores the inadequacy of a simple superposition principle; a unified theory like NEGF is essential to capture the interplay where one transport channel actively modifies the other.

\begin{figure}
	\centering
	\includegraphics[width=1.0\textwidth]{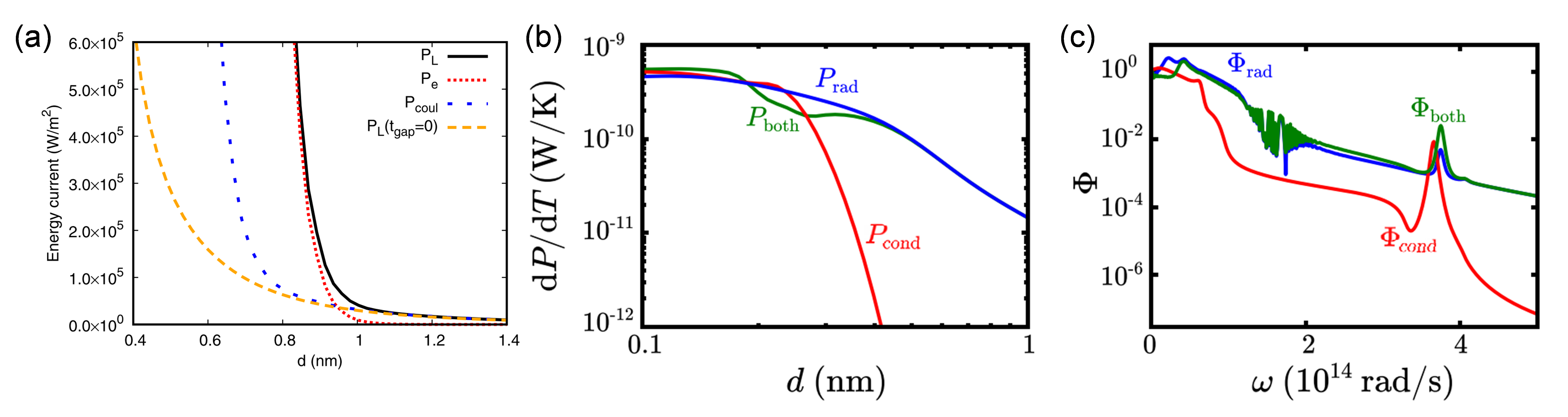}
	\caption{(a) Photon-Electron Synergy: Heat flux in a metal-vacuum-metal junction as a function of distance. The total heat flux (solid black line) is a non-trivial combination of contributions from Coulomb fluctuations (photons, blue dashed line) and electron tunneling (electrons, red dotted line). A crossover occurs at $d \approx0.92$ nm, below which electron tunneling dominates. In the strong tunneling regime ($d < 0.6$ nm), the total flux exceeds the simple sum of the two, indicating non-linear enhancement. (b) Photon-Phonon Synergy: Thermal conductance between two collinear carbyne wires. The total conductance including both radiative (RHT) and phonon (PCHT) channels (solid green line, ``both") is shown along with the individual contributions. In the range $d \approx0.2-0.4$ nm, the total conductance is anomalously suppressed below either of the individual channels, demonstrating their non-additive nature. (c) Spectral Origin of Suppression: The Landauer energy transmission spectrum for the carbyne wires at $d = 0.281$ nm. The coupled spectrum (``both") is significantly suppressed in the low-frequency region compared to the pure radiative spectrum (``rad"), explaining the reduction in total conductance seen in (b). Panel (a) adapted with permission from~\cite{add62}. Copyright 2018 American Physical Society. Panels (b) and (c) adapted with permission from~\cite{add33}. Copyright 2020 American Physical Society.}
	\label{fig2}
\end{figure}

A similar non-additive synergy exists between radiative heat transfer (photons) and phonon conduction heat transfer (PCHT). Venkataram \textit{et al}.~\cite{add33} developed a generalized NEGF formalism for linear bosons to provide a common mechanical basis for both RHT and PCHT, applying it to the case of two collinear carbyne wires. Their results revealed a striking and counter-intuitive phenomenon: in the distance range of $d\approx 0.2 - 0.4$ nm, the total thermal conductance is anomalously suppressed, falling below the conductance from either the radiative or the conductive channel alone (Fig.~\ref{fig2}b). This anomalous suppression arises from a destructive interference effect; the electromagnetic coupling between the vibrating atoms in the two wires alters the system's resonant frequencies and suppresses the response amplitude of both photons and phonons (Fig.~\ref{fig2}c). Analysis of the Landauer energy transmission spectrum confirms that the coupling of the two mechanisms significantly reduces the heat transfer contribution in the critical low-frequency range. This demonstrates that at the atomic scale, treating different heat transfer mechanisms as independent parallel pathways is not just an approximation but can be qualitatively wrong.

These studies showcase the unique power of NEGF to unify the description of heat transfer. By treating photons, electrons, and phonons within a single, coherent quantum framework, NEGF moves beyond the classical paradigm of summing independent contributions. It reveals a richer, more complex reality of synergistic and competitive transport at the quantum limit, a regime where such a unified theory is not just advantageous, but essential.

\begin{figure}
	\centering
	\includegraphics[width=1.0\textwidth]{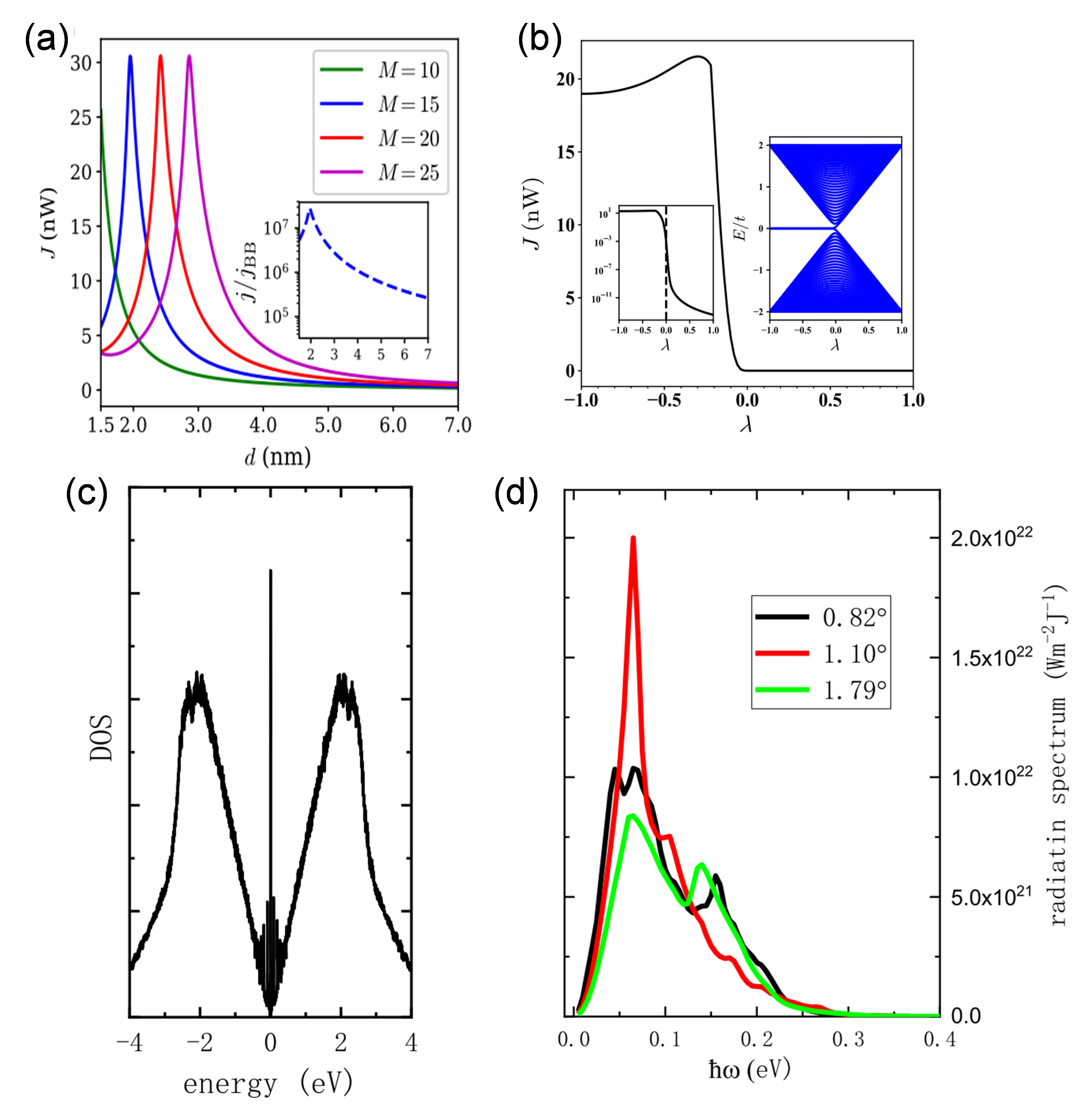}
	\caption{(a) Topological Control: Heat flux between two zigzag carbon nanotubes exhibits a non-monotonic dependence on distance, with a peak at a critical distance $d_c$. This anomalous behavior is a direct consequence of topologically protected edge states. (b) Topological Thermal Switch: By tuning a hopping parameter $\lambda$ in an SSH chain model to induce a topological phase transition (inset shows the emergence of a zero-energy edge state), the heat flux can be switched on and off. (c) Band Structure Engineering: The density of states (DOS) of twisted bilayer graphene (TBG) exhibits sharp van Hove singularities near the Fermi level when tuned to the magic angle ($\theta = 1.1^\circ$), a result of engineered flat bands. (d) Tunable Thermal Emission: The engineered flat bands in TBG lead to a sharp, intense peak in its thermal radiation spectrum. The position of this peak is highly tunable with the twist angle, allowing the material's radiative properties to be designed for specific functions. Panels (a) and (b) adapted with permission from~\cite{add86}. Copyright 2019 American Physical Society. Panels (c) and (d) adapted with permission from~\cite{add93}. Copyright 2022 Elsevier.}
	\label{fig3}
\end{figure}
\subsection{From Materials to Metamaterials: Engineering RHT via Quantum Design}

Beyond providing a more accurate description of existing materials, the true predictive power of the NEGF formalism lies in its ability to connect the fundamental quantum-mechanical design of a system to its macroscopic thermal radiation properties. This capability elevates NEGF from a mere analysis tool to a design framework, enabling the engineering of materials and metamaterials with bespoke thermal functionalities. By manipulating the underlying electronic Hamiltonian---through topology, geometry, or Moir\'{e} stacking---one can proactively control heat flow, a paradigm shift from passively characterizing it.

A compelling demonstration of this principle is the control of NFRHT via ``topological engineering". The intrinsic topology of a material's band structure can mandate the existence of protected edge states, which dramatically alter its transport properties. NEGF calculations by Tang \textit{et al}~\cite{add86} have revealed the profound impact of such zero-energy edge states on NFRHT in systems like zigzag carbon nanotubes and graphene nanotriangles. The presence of these highly localized states near the Fermi level creates a powerful, low-energy channel for heat transfer~\cite{add87,add88,add89}. This leads not only to a giant enhancement of heat flux---up to seven orders of magnitude above the blackbody limit---but also to a highly anomalous, non-monotonic dependence on the vacuum gap distance, where the flux is maximized at an optimal critical distance, $d_c$ (Fig.~\ref{fig3}a). This behavior stands in stark contrast to the monotonic power-law decay seen in non-topological systems. The physical mechanism, clearly elucidated by NEGF through an analysis of the spectral transmission function, involves resonant tunneling mediated by the coupled edge states. Most importantly, this topological effect can be harnessed for active control (Fig.~\ref{fig3}b). As demonstrated with a Su-Schrieffer-Heeger (SSH) model~\cite{add90}, by tuning a parameter that drives the system through a topological phase transition, the edge states can be switched ``on" or ``off". This translates directly into a thermal switching effect, providing a blueprint for a topological thermal switch.

The concept of quantum design extends beyond topology to direct ``band structure engineering" via Moir\'{e} physics~\cite{add91}. Twisted bilayer graphene (TBG) has emerged as the archetypal ``twistronic" metamaterial, where the twist angle acts as a continuous tuning parameter for its electronic properties~\cite{add92}. NEGF-based calculations have been pivotal in connecting the engineered band structure of TBG to its thermal radiation signature~\cite{add93,add94}. Near the ``magic angle" ($\theta\approx 1.1^\circ$), strong interlayer coupling flattens the electronic bands, creating sharp van Hove singularities in the density of states near the Fermi level (Fig.~\ref{fig3}c)~\cite{add93,add95}. This engineered electronic structure drastically alters the material's optical response~\cite{add91}. The flat bands open up strong, low-energy interband transitions, resulting in a sharp and intense peak in TBG's far-field thermal emission spectrum (Fig.~\ref{fig3}d). This feature is highly tunable: by changing the twist angle, the position of the emission peak can be precisely controlled. This allows for the design of materials with tailored radiative properties; for instance, the emission peak can be positioned outside the atmospheric transparency window for applications in thermal insulation or infrared stealth, or shifted into the window to switch its function.

In both examples, NEGF serves as the essential theoretical bridge, translating an abstract change in the quantum Hamiltonian---be it topological or geometric---into a concrete, predictable, and often dramatic change in macroscopic heat transfer. This opens a new frontier where the principles of quantum design can be used to forge novel materials and metamaterials with on-demand thermal radiation characteristics.

\subsection{Active Control: Driving Heat Flow in Nonequilibrium Systems}
\begin{figure}
	\centering
	\includegraphics[width=1.0\textwidth]{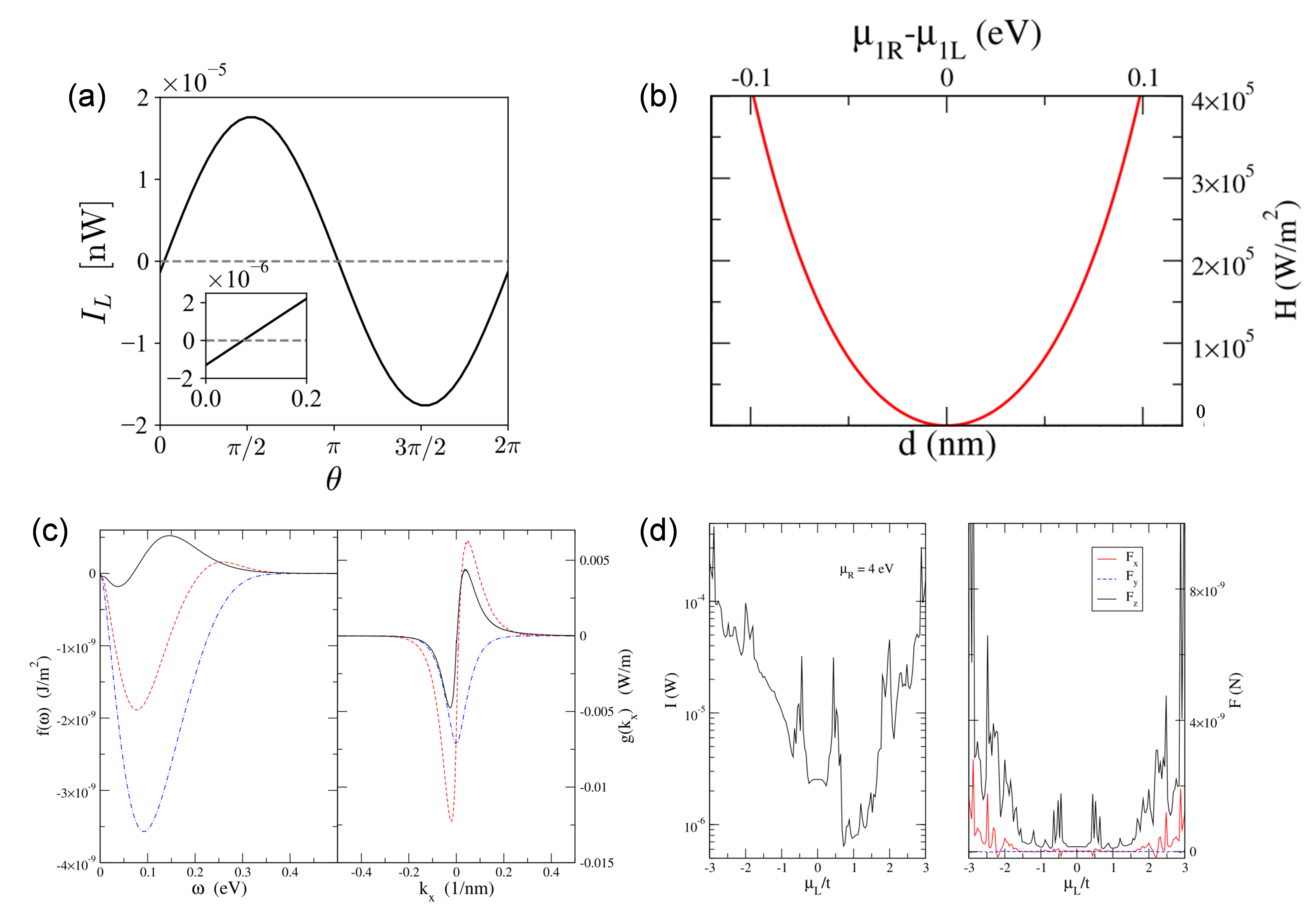}
	\caption{(a) Floquet Phase Control: In a system of two bodies held at the same temperature ($T_1 = T_2$), a net energy flux can be generated by driving them with a relative phase difference $\theta = \theta_L-\theta_R$. The plot shows that the heat flux is directly controlled by this phase, enabling its magnitude and direction to be tuned externally. This provides a mechanism for thermal rectification. (b) Current-Induced Heat Flux: Heat flux between two graphene layers at the same temperature as a function of distance, induced solely by a drift current in one layer. The flux exhibits a parabolic dependence on the drift velocity $v_d$, demonstrating that the external drive is the source of the energy transfer.  (c) Current-Driven Thermal Shutoff: The plot shows the spectral heat flux between two graphene layers with a temperature difference. Without a current (blue dash-dotted line), heat flows from hot to cold. An applied drift current induces a negative heat flux component via negative Landau damping (red dashed line). A sufficiently strong current (black solid line) can pump heat against the gradient, achieving a ``thermal shutoff" or even active cooling. (d) Current-Induced Heat Transfer and Forces: In an asymmetrically biased graphene nanoribbon-$C_{60}$ system, the non-equilibrium current generates not only heat transfer (left panel) but also measurable forces (right panel) on the nanoribbon, with sharp resonant peaks appearing at specific bias voltages. Panel (a) adapted with permission from~\cite{add24}. Copyright 2024 American Physical Society. Panels (b) and (c) adapted with permission from a preprint by Peng \textit{et al}~\cite{add31}. Permission to reuse was requested from the authors. Panel (d) adapted with permission from~\cite{add30}. Copyright 2024 American Physical Society.}
	\label{fig4}
\end{figure}
The most profound departure from classical thermal science, and arguably the most significant frontier opened by NEGF, is the study of RHT in genuine non-equilibrium steady states. In traditional scenarios, heat flow is a passive process, driven by a pre-existing temperature gradient. However, by externally driving a system---for instance, with time-varying fields or a DC electrical current---it is possible to actively control and manipulate the flow of thermal energy. In these driven systems, the material's constituents are no longer in local thermal equilibrium, and the FDT, the cornerstone of conventional FE theory, fundamentally fails~\cite{add25,add26,add27}. NEGF, which is constructed from the ground up to handle quantum transport far from equilibrium, provides the natural and necessary language to describe this new physics.

Floquet engineering, where a system parameter is modulated periodically in time, offers a powerful route to active thermal control. Such periodic driving shatters the system's time-translation symmetry, forcing it into a non-equilibrium Floquet steady state~\cite{add96,add97}. NEGF calculations have shown that this process can populate electromagnetic modes in a non-thermal manner, creating an effective photon distribution that corresponds to a temperature much higher than the ambient temperature of the system itself. The most striking consequence of this is the ability to induce a net heat flux between two bodies held at the ``identical" ambient temperature. This seemingly paradoxical ``isothermal heat transfer", forbidden in equilibrium, becomes possible because the external drive acts as an energy pump~\cite{add28}. By introducing an asymmetry---either by using two different materials or by driving two identical materials with a relative phase difference, a directional heat flux can be generated~\cite{add24}. As shown in Fig.~\ref{fig4}a, the magnitude and even the direction of this heat flux can be precisely controlled by the driving phase, providing a blueprint for designing novel devices like near-field thermal rectifiers and nonreciprocal heat switches. Furthermore, Floquet modulation can be used to overcome the intrinsic limitations of NFRHT by actively converting the energy of non-radiative evanescent waves into propagating far-field radiation, enabling efficient near-field to far-field energy transfer~\cite{add29}.

An alternative path to non-equilibrium control is through ``current driving", where a DC voltage bias creates a steady-state drift current of electrons. This directed motion breaks the system's local equilibrium and enables a new mechanism for photon emission: ``negative Landau damping"~\cite{add98}. When the electron drift velocity exceeds the phase velocity of an electromagnetic mode, the electrons can amplify the mode, leading to spontaneous photon emission even in the absence of a thermal gradient. NEGF theory has been instrumental in exploring the consequences of this effect. As shown in Fig.~\ref{fig4}b, it predicts that a drift current in one of two parallel graphene layers can generate a significant radiative heat flux even when both layers are at the same temperature~\cite{add31}. More remarkably, a sufficiently strong drift current can pump heat against a temperature gradient, transferring energy from a colder body to a hotter one. This leads to the possibility of active cooling or achieving a ``thermal shutoff", where the current-induced heat flux precisely cancels the natural heat flow from the temperature gradient (Fig.~\ref{fig4}c). Beyond energy transfer, NEGF calculations also predict that these non-equilibrium currents can induce optical forces and torques on nearby objects (Fig.~\ref{fig4}d), opening avenues for coupled energy and momentum manipulation at the nanoscale~\cite{add30}.

In both Floquet and current-driven scenarios, NEGF provides the essential framework to go beyond equilibrium statistical mechanics. It transforms the concept of heat transfer from a passive consequence of temperature differences into an actively controllable process, governed by external driving parameters like phase, frequency, and voltage. This paradigm shift paves the way for a new generation of active thermal management technologies, energy conversion devices, and thermal information processing systems.

\section{Outlook and Future Perspectives}
We have shown that NEGF-based studies have already provided many new insights on RHT. However, the application of NEGF to radiative heat transfer is still in its early stages, and many exciting directions lie ahead. Here, we outline some future perspectives and challenges in this emerging field:

1. Unified Nano-Heat-Transfer Framework: The ultimate promise of NEGF in this context is a unified theory of heat exchange that seamlessly treats conduction, radiation, and even convection (if one includes particle exchange) within one framework. Real nanoscale systems --- such as transistor circuits, optical chips, or MEMS devices --- often involve multiple heat pathways. A single formalism could handle overlapping regimes (for instance, a pair of surfaces that are close enough for electrons to tunnel as well as radiative tunneling via photons). While recent work has initiated this unification~\cite{add33,add62}, a critical challenge remains in describing RHT between polar materials, where phonon-photon coupling mediates the transfer. The long-range nature of dipolar interactions in these systems poses significant computational challenges for direct NEGF calculations, requiring manipulation of extremely large and dense matrices. Recent advances using atomistic methods have successfully studied extreme near-field heat transfer between polar materials~\cite{add99,add100,add101}. However, these approaches typically specialize in the extreme near-field regime. A major future challenge therefore involves developing a unified NEGF formalism capable of seamlessly describing RHT in polar materials across all length scales—from the extreme near-field to the far-field—while maintaining computational tractability.

2. \textit{Ab Initio} Material Descriptions: Both NEGF and FE depend on modeling materials' self-energy or response functions to electromagnetic fields. Incorporating realistic electronic band structures and vibrational spectra from density functional theory (DFT) or other first-principles methods into NEGF radiative transfer calculations represents a promising direction. Progress in first-principles quantum transport indicates that developing a first-principles NEGF for photons is achievable~\cite{add71,add72}. Such efforts will improve the quantitative predictive power of near-field RHT, especially in extreme conditions (sub-nm gaps, molecular junctions) where current FE predictions might be unreliable. They will also clarify how material properties at the atomistic level (e.g. band gap, electron density, phonon dispersions) govern the radiative vs. conductive heat partition. Again, a potential challenge here is the heavy computational cost --- NEGF calculations with rich atomistic detail are computationally demanding, and adding the electromagnetic coupling increases complexity. Advances in computational algorithms and power, perhaps combined with machine-learning surrogates for Green's functions, may be needed to tackle realistic setups.

3. Many-Body and Nonlinear Effects: Thus far, most NEGF studies of RHT have considered either two-body systems or linear responses. An open question is how to handle many-body radiative heat transfer (multiple objects exchanging heat simultaneously) in a NEGF framework. Fluctuational electrodynamics has a scattering theory extension to many bodies, but its complexity grows quickly. NEGF, on the other hand, is conceptually well-suited to network or circuit geometries --- in principle one can include multiple contacts or reservoirs. Developing a NEGF network theory for radiative heat (analogous to multi-terminal electric circuits) could address heat flow in nanoscale networks, e.g. an array of nanoparticles or a complex quantum thermal device. Additionally, NEGF could be extended to examine nonlinear regimes of radiative transfer. For example, if temperature differences are very large or if one intentionally drives a system strongly (far beyond linear response), the radiative exchange might become nonlinear in temperature or exhibit saturation effects. Including higher-order interactions (beyond RPA, for instance) in NEGF could capture phenomena like thermal rectification (asymmetric heat flow) or photon bunching effects in intense thermal fields. Some initial studies have hinted at thermal rectification via nonlinear feedback~\cite{add102}, but a rigorous NEGF treatment is yet to be done. Nonlinear NEGF for photons could also shed light on regimes where material responses themselves become temperature-dependent during exchange, leading to feedback.

4. Dynamic Modulation and Thermal Signal Processing: The Floquet study discussed above hints at the broader opportunity to control radiative heat transfer dynamically~\cite{add24,add28,add29}. NEGF is naturally suited to treat time-dependent transport, so one can study scenarios like modulating a gap spacing or an optical property in time to achieve active control of heat flow. For instance, a rapidly time-varying mirror could selectively suppress or enhance certain frequency components of thermal radiation. Such dynamic radiative thermal management could enable thermally switchable materials, thermal logic gates, or novel cooling strategies. One vision is a thermal transistor where a gate input (perhaps light or voltage) modulates the radiative heat flow between two terminals. FE theory can handle some modulation in a quasi-static way, but NEGF would allow studying truly high-frequency modulation and non-equilibrium states thereby created. Moreover, by capturing the spectral distribution of non-equilibrium photons, NEGF could guide the design of thermophotovoltaic devices and energy conversion systems that operate under non-ideal conditions (for example, a hot emitter out of equilibrium with a photonic band excitation --- something beyond a simple blackbody). We anticipate that developing intuitive circuit models for radiative heat flow (where photons, phonons, electrons are analogous to currents through thermal resistances) will be an outcome of this line of research, aiding engineers in designing complex thermal networks.

5. Quantum and Information-Theoretic Aspects: As we push radiative heat transfer into regimes involving few quanta or requiring quantum coherence (e.g. at very low temperatures, or between quantum cavities), NEGF might play a role in connecting to quantum information theory. For example, one could ask: can entangled or squeezed light be used to enhance or suppress heat transfer? Questions about the quantum limit of thermal photonic conductance, or the entropy carried by radiative heat flow, are largely unexplored. NEGF provides access to the quantum statistical properties of the radiation field (via higher-order Green's functions or correlators), so it might help quantify entropic and information content of heat flow at the nanoscale. While these ideas are somewhat speculative, the framework could potentially incorporate quantum state engineering into thermal radiation problems --- a realm completely untouched by classical FE.

Finally, a practical perspective: with continued advances in nanofabrication and experimental thermal metrology, some of the exotic predictions of NEGF-based RHT may become testable. For instance, one could envision experiments to verify heat transfer with zero temperature difference using modulated materials (as in the Floquet scenario), or to measure the noise (fluctuations) in near-field heat currents to confirm the predicted $2k_B T^2$ relation in extreme near-field settings~\cite{add103}. Such experiments would not only validate the NEGF approach but also potentially lead to novel thermal technologies (for example, new ways to refrigerate using only radiation and no temperature gradient, by exploiting non-equilibrium distributions). The synergy between theory and experiment will be crucial: NEGF provides a rich, nuanced picture of thermal radiation at the nanoscale, and experiments will both test these predictions and inspire refinements to the theory.

\section{Conclusion}
In summary, the recent application of non-equilibrium Green's function methods to radiative heat transfer represents a paradigm shift in how we analyze and understand thermal radiation at the nanoscale. Building on decades of success with fluctuational electrodynamics, the NEGF approach broadens the scope to genuine non-equilibrium situations, unifies radiation with conduction, and lays bare the quantum statistical underpinnings of heat exchange. We have reviewed how NEGF-based studies have already provided new insights --- from the dominance of electron Coulomb interactions in extreme near-field heat transfer, to the non-additivity of radiative and conductive heat at the atomic scale, to the possibility of radiative heat flow without a temperature difference by external modulation. These developments not only deepen our fundamental understanding but also open routes to novel thermal devices and control strategies.

As this burgeoning field moves forward, we expect rapid progress on both theoretical and experimental fronts. The NEGF formalism will likely be further refined and integrated with first-principles calculations to improve quantitative accuracy. Its predictions, particularly in novel regimes (e.g. time-modulated or quantum-coherent thermal radiation), will stimulate experiments that test the limits of thermal science. In parallel, concepts from NEGF could be distilled into effective models or simplified ``thermal circuit" analogues for use by engineers in advanced thermal management applications. The cross-disciplinary nature of this research --- intersecting condensed matter physics, thermodynamics, and nanotechnology --- is poised to yield a rich understanding of heat in the modern world. The future perspective is clear: by embracing non-equilibrium approaches like NEGF, we will unravel new regimes of radiative heat transfer and harness them in ways not possible before, continuing the cycle of innovation in the science of energy transfer at the smallest scales.

%
%

\ack{This work is supported by the National Natural Science Foundation of China (Grant No. 12204346) and the National Key R$\&$D program of China (Project No. 2022YFA1204000).}



\data{All data that support the findings of this study are included within the article (and any supplementary files).}


\end{justify}

\begin{thebibliography}{}
    \bibitem{add1} Rousseau E, Siria A, Jourdan G, Volz S, Comin F, Chevrier J and Greffet J-J 2009 Radiative heat transfer at the nanoscale \textit{Nat. Photonics} \textbf{3} 514-7
    \bibitem{add2} Howell J R, Meng\"{u}c M P, Daun K and Siegel R 2020 \textit{Thermal Radiation Heat Transfer} (Boca Raton: CRC Press)
    \bibitem{add3} Modest M F 2003 \textit{Radiative Heat Transfer} (Burlington: Academic Press)
    \bibitem{add4} Planck M 1989 \textit{The Theory of Heat Radiation} (American Institute of Physics Melville, NY)
    \bibitem{add5} Kittel A, M\"{u}ller-Hirsch W, Parisi J, Biehs S-A, Reddig D and Holthaus M 2005 Near-field heat transfer in a scanning thermal microscope \textit{Phys. Rev. Lett.} \textbf{95} 224301
	\bibitem{add6} Basu S, Zhang Z M and Fu C J 2009 Review of near-field thermal radiation and its application to energy conversion \textit{Int. J. Energy Res.} \textbf{33} 1203-32
	\bibitem{add7} Song B, Fiorino A, Meyhofer E and Reddy P 2015 Near-field radiative thermal transport: From theory to experiment \textit{AIP Adv.} \textbf{5} 053503
    \bibitem{add8} Joulain K, Mulet J-P, Marquier F, Carminati R and Greffet J-J 2005 Surface electromagnetic waves thermally excited: Radiative heat transfer, coherence properties and Casimir forces revisited in the near field \textit{Surf. Sci. Rep.} \textbf{57} 59-112
	\bibitem{add9} Volokitin A I and Persson B N J 2007 Near-field radiative heat transfer and noncontact friction \textit{Rev. Mod. Phys.} \textbf{79} 1291-329
	\bibitem{add10} Biehs S-A, Messina R, Venkataram P S, Rodriguez A W, Cuevas J C and Ben-Abdallah P 2021 Near-field radiative heat transfer in many-body systems \textit{Rev. Mod. Phys.} \textbf{93} 025009
	\bibitem{add11} Henkel C 2017 Nanoscale thermal transfer -- An invitation to fluctuation electrodynamics \textit{Z. F\"{u}r Naturforschung A} \textbf{72} 99-108
	\bibitem{add12} Pascale M, Giteau M and Papadakis G T 2023 Perspective on near-field radiative heat transfer \textit{Appl. Phys. Lett.} \textbf{122} 100501
	\bibitem{add13} Rytov S M, 1953 \textit{Theory of Electrical Fluctuation and Thermal Radiation} (Academy of Science of USSR, Moscow)
    \bibitem{add14} Volokitin A I and Persson B N J 2017 Theory of the Fluctuating Electromagnetic Field \textit{Electromagnetic Fluctuations at the Nanoscale: Theory and Applications} (Berlin, Heidelberg: Springer)
	\bibitem{add15} Rytov S M, Kravtsov Y A and Tatarskii V I 1989 \textit{Principles of Statistical Radiophysics 3} (Springer Berlin, Heidelberg)
    \bibitem{add16} Callen H B and Welton T A 1951 Irreversibility and generalized noise \textit{Phys. Rev.} \textbf{83} 34-40
    \bibitem{add17} Polder H B and Van Hove M 1971 Theory of radiative heat transfer between closely spaced bodies \textit{Phys. Rev. B} \textbf{4}, 3303.
    \bibitem{add18} Otey C R, Lau W T and Fan S 2010 Thermal rectification through vacuum \textit{Phys. Rev. Lett.} \textbf{104} 154301
    \bibitem{add19} Kr\"{u}ger M 2011 Nonequilibrium electromagnetic fluctuations: heat transfer and interactions \textit{Phys. Rev. Lett.} \textbf{106} 210404
    \bibitem{add20} Tang G, Zhang L, Zhang Y, Chen J and Chan C T 2021 Near-Field energy transfer between graphene and magneto-optic media \textit{Phys. Rev. Lett.} \textbf{127} 247401
	\bibitem{add21} Ottens R S, Quetschke V, Wise S, Alemi A A, Lundock R, Mueller G, Reitze D H, Tanner D B and Whiting B F 2011 Near-field radiative heat transfer between macroscopic planar surfaces \textit{Phys. Rev. Lett.} \textbf{107} 014301
	\bibitem{add22} Kim K, Song B, Fern\'{a}ndez-Hurtado V, Lee W, Jeong W, Cui L, Thompson D, Feist J, Reid M T H, Garc\'{i}a-Vidal F J, Cuevas J C, Meyhofer E and Reddy P 2015 Radiative heat transfer in the extreme near field \textit{Nature} \textbf{528} 387-91
    \bibitem{add23} Shen S, Narayanaswamy A and Chen G 2009 Surface phonon polaritons mediated energy transfer between nanoscale gaps \textit{Nano Lett.} \textbf{9} 2909-13
    \bibitem{add24} Tang G and Wang J-S 2024 Modulating near-field thermal transfer through temporal drivings: A quantum many-body theory \textit{Phys. Rev. B} \textbf{109} 085428
    \bibitem{add25}  V\'{a}zquez-Lozano J E and Liberal I 2023 Incandescent temporal metamaterials \textit{Nat. Commun.} \textbf{14} 4606
	\bibitem{add26} Aoki H, Tsuji N, Eckstein M, Kollar M, Oka T and Werner P 2014 Nonequilibrium dynamical mean-field theory and its applications \textit{Rev. Mod. Phys.} \textbf{86} 779-837
	\bibitem{add27} Kohn W 2001 Periodic thermodynamics \textit{J. Stat. Phys.} \textbf{103} 417-23
    \bibitem{add28} Pan H, Ren Y, Tang G and Wang J-S 2025 Asymmetry-induced radiative heat transfer in Floquet systems \textit{Phys. Rev. B} \textbf{112} L041401
    \bibitem{add29} Zhu H, Ren Y, Pan H, Tang G, Zhang L and Wang J-S 2025 arXiv:2507.16688
    \bibitem{add30} Wang J-S and Antezza M 2024 Photon mediated energy, linear and angular momentum transport in fullerene and graphene systems beyond local equilibrium \textit{Phys. Rev. B} \textbf{109} 125105
	\bibitem{add31} Peng J and Wang J-S 2019 arXiv:1805.09493v2
    \bibitem{add32} Zhang Y-M, Zhu T, Zhang Z-Q and Wang J-S 2022 Microscopic theory of photon-induced energy, momentum, and angular momentum transport in the nonequilibrium regime \textit{Phys. Rev. B} \textbf{105} 205421
    \bibitem{add33} Venkataram P S, Messina R, Cuevas J C, Ben-Abdallah P and Rodriguez A W 2020 Mechanical relations between conductive and radiative heat transfer \textit{Phys. Rev. B} \textbf{102} 085404
    \bibitem{add34} Wang J-S, Peng J, Zhang Z-Q, Zhang Y-M and Zhu T 2023 Transport in electron-photon systems \textit{Front. Phys.} \textbf{18} 43602
    \bibitem{add35} Wang J-S, Agarwalla B K, Li H and Thingna J 2014 Nonequilibrium Green's function method for quantum thermal transport \textit{Front. Phys.} \textbf{9} 673-97
	\bibitem{add36} Zhu T, Zhang Y-M and Wang J-S 2024 Super-Planckian radiative heat transfer between coplanar two-dimensional metals \textit{Phys. Rev. B} \textbf{109} 245427
	\bibitem{add37} Keldysh L V 1965 Diagram technique for nonequilibrium processes \textit{Sov Phys JETP} \textbf{20} 1018-26
	\bibitem{add38} Haug H and Jauho A-P 2008 \textit{Quantum Kinetics in Transport and Optics of Semiconductors} vol 123 (Berlin, Heidelberg: Springer)
	\bibitem{add39} Wang J-S, Wang J and L\"{u} J T 2008 Quantum thermal transport in nanostructures \textit{Eur. Phys. J. B} \textbf{62} 381-404
	\bibitem{add40} Di Ventra M 2008 \textit{Electrical Transport in Nanoscale Systems} (Cambridge University Press)
	\bibitem{add41} Datta S 1995 \textit{Electronic Transport in Mesoscopic Systems} (Cambridge University Press)
    \bibitem{add42} Magpantay J A 1994 The Coulomb Gauge Revisited \textit{Prog. Theor. Phys.} \textbf{91} 573-7
    \bibitem{add43} Creutz M 1979 Quantum electrodynamics in the temporal gauge \textit{Ann. Phys.} \textbf{117} 471-83
	\bibitem{add44} Heisenberg W and Pauli W 1930 Zur quantentheorie der wellenfelder. II \textit{Z. F\"{u}r Phys.} \textbf{59} 168-90
    \bibitem{add45} Weinberg S 1995 \textit{The Quantum Theory of Fields} vol 1 (Cambridge University Press)
    \bibitem{add46} Graf M and Vogl P 1995 Electromagnetic fields and dielectric response in empirical tight-binding theory \textit{Phys. Rev. B} \textbf{51} 4940-9
	\bibitem{add47} Peierls R 1933 Zur theorie des diamagnetismus von Leitungselektronen \textit{Z. F\"{u}r Phys.} \textbf{80} 763-91
    \bibitem{add48} Fetter A L and Walecka J D 1971 \textit{Quantum Theory of Many-Particle Systems} (McGraw-Hill)
	\bibitem{add49} Ford G W, Kac M and Mazur P 1965 Statistical mechanics of assemblies of coupled oscillators \textit{J. Math. Phys.} \textbf{6} 504-15
	\bibitem{add50} Dyson F J 1949 The S Matrix in Quantum Electrodynamics \textit{Phys. Rev.} \textbf{75} 1736
	\bibitem{add51} Langreth D C 1976 \textit{Linear and Nonlinear Electron Transport in
	Solids} (Springer, Boston, 1976)
    \bibitem{add52} Keller O 2011 \textit{Quantum Theory of Near-Field Electrodynamics} (Berlin, Heidelberg: Springer)
	\bibitem{add53} Novotny L and Hecht B 2012 \textit{Principles of Nano-Optics} (Cambridge University Press)
    \bibitem{add54} Rammer J 2007 \textit{Many-Body Quantum Theory in Condensed Matter Physics: An Introduction} (Cambridge University Press)
	\bibitem{add55} Rammer J 2007 \textit{Quantum Field Theory of Non-equilibrium States} (Cambridge University Press)
	\bibitem{add56} Ehrenreich H and Cohen M H 1959 Self-Consistent Field Approach to the Many-Electron Problem \textit{Phys. Rev.} \textbf{115} 786
	\bibitem{add57} Jackson J D 2009 \textit{Classical electrodynamics} (Hoboken, NY: Wiley)
	\bibitem{add58} Meir Y and Wingreen N S 1992 Landauer formula for the current through an interacting electron region \textit{Phys. Rev. Lett.} \textbf{68} 2512.
	\bibitem{add59} Jauho A-P, Wingreen N S and Meir Y 1994 Time-dependent transport in interacting and noninteracting resonant-tunneling systems \textit{Phys. Rev. B} \textbf{50} 5528
	\bibitem{add60} Caroli C, Combescot R, Nozieres P and Saint-James D 1971 Direct calculation of the tunneling current \textit{J. Phys. C} \textbf{4} 916
	\bibitem{add61} Jiang J-H and Wang J-S 2017 Caroli formalism in near-field heat transfer between parallel graphene sheets \textit{Phys. Rev. B} \textbf{96} 155437
    \bibitem{add62} Zhang Z-Q, L\"{u} J-T and Wang J-S 2018 Energy transfer between two vacuum-gapped metal plates: Coulomb fluctuations and electron tunneling \textit{Phys. Rev. B} \textbf{97} 195450
	\bibitem{add63} Zhu T and Wang J-S 2021 Generalized first-principles method to study near-field heat transfer mediated by Coulomb interaction \textit{Phys. Rev. B} \textbf{104} L121409
	\bibitem{add64} Ilic O, Jablan M, Joannopoulos J D, Celanovic I, Buljan H and Solja\u{c}i\'{c} M 2012 Near-field thermal radiation transfer controlled by plasmons in graphene \textit{Phys. Rev. B} \textbf{85} 155422
    \bibitem{add65} Parr R G 1980 \textit{Density Functional Theory of Atoms and Molecules Horizons of Quantum Chemistry} (Dordrecht: Springer Netherlands)
    \bibitem{add66} Wang J-S, Zhang Z-Q and L\"{u} J-T 2018 Coulomb-force-mediated heat transfer in the near field: Geometric effect \textit{Phys. Rev. E} \textbf{98} 012118
    \bibitem{add67} Peng J, Yap H H, Zhang G and Wang J-S 2017 arXiv:1703.07113
	\bibitem{add68} Wang J-S and Peng J 2017 Capacitor physics in ultra-near-field heat transfer \textit{EPL Europhys. Lett.} \textbf{118} 24001
    \bibitem{add69} Mahan G D 2017 Tunneling of heat between metals, \textit{Phys. Rev. B} \textbf{95} 115427
    \bibitem{add70}Paulsson M, Frederiksen T and Brandbyge M 2005 Modeling inelastic phonon scattering in atomic- and molecular-wire junctions \textit{Phys. Rev. B} \textbf{72} 201101
	\bibitem{add71} Zhu T, Zhang Z-Q, Gao Z and Wang J-S 2020 First-principles method to study near-field radiative heat transfer \textit{Phys. Rev. Appl.} \textbf{14} 024080
	\bibitem{add72} Zhu T, Trevisanutto P E, Asmara T C, Xu L, Feng Y P and Rusydi A 2018 Generation of multiple plasmons in strontium niobates mediated by local field effects \textit{Phys. Rev. B} \textbf{98} 235115
	\bibitem{add73} Adler S L 1962 Quantum theory of the dielectric constant in real solids \textit{Phys. Rev.} \textbf{126} 413-20
	\bibitem{add74} Wiser N 1963 Dielectric constant with local field effects included \textit{Phys. Rev.} \textbf{129} 62-9
    \bibitem{add75} Onida G, Reining L and Rubio 2002 A Electronic excitations: density-functional versus many-body Green's-function approaches \textit{Rev. Mod. Phys.} \textbf{74} 601
    \bibitem{add76} Zhu T, Zheng C, Xu L and Yang M 2024 Exciton dissociation in two-dimensional transition metal dichalcogenides: Excited states and substrate effects \textit{Phys. Rev. B} \textbf{110} 155416
    \bibitem{add77} Salpeter E E and Bethe H A 1951 \textit{Phys. Rev.} A Relativistic Equation for Bound-State Problems \textbf{84} 1232
    \bibitem{add78} Zhu T, Antezza M and Wang J-S 2021 Dynamical polarizability of graphene with spatial dispersion \textit{Phys. Rev. B} \textbf{103} 125421
    \bibitem{add79} Kittel A, M\"{u}ller-Hirsch W, Parisi J, Biehs S-A, Reddig D 2005 Near-Field Heat Transfer in a Scanning Thermal Microscope \textit{Phys. Rev. Lett.} \textbf{95} 224301
    \bibitem{add80} Chapuis P-O, Volz S, Henkel C, Joulain K and Greffet J-J 2008 Effects of spatial dispersion in near-field radiative heat transfer between two parallel metallic surfaces \textit{Phys. Rev. B} \textbf{77} 035431
    \bibitem{add81} Loomis J J and Maris H J 1994 Theory of heat transfer by evanescent electromagnetic waves \textit{Phys. Rev. B} \textbf{50} 18517-24
	\bibitem{add82} Grigorenko A N, Polini M and Novoselov K S 2012 Graphene plasmonics \textit{Nat. Photonics} \textbf{6} 749-58
    \bibitem{add83} Wang J-S 2025 Beyond the Drude model: Surface and non-local effects in near-field radiative heat transfer and the Casimir puzzle \textit{Phys. Rev. B} \textbf{111} 245404
	\bibitem{add84} Lian K N and Wang J-S 2020 Geometric effect on near-field heat transfer analysis using efficient graphene and nanotube models \textit{Eur. Phys. J. B} \textbf{93} 138
    \bibitem{add85} Cui L, Jeong W, Fern\'{a}ndez-Hurtado V, Feist J, Garc\'{i}a-Vidal F J, Cuevas J C, Meyhofer E and Reddy P 2017 Study of radiative heat transfer in \AA ngstr\"{o}m- and nanometre-sized gaps \textit{Nat. Commun.} \textbf{8} 14479
    \bibitem{add86} Tang G, Yap H H, Ren J and Wang J-S 2019 Anomalous near-field heat transfer in carbon-based nanostructures with edge states \textit{Phys. Rev. Appl.} \textbf{11} 031004
	\bibitem{add87} Sasaki K, Sato K, Saito R, Jiang J, Onari S and Tanaka Y 2007 Local density of states at zigzag edges of carbon nanotubes and graphene \textit{Phys. Rev. B} \textbf{75} 235430
	\bibitem{add88} Nakada K, Fujita M, Dresselhaus G and Dresselhaus M S 1996 Edge state in graphene ribbons: Nanometer size effect and edge shape dependence \textit{Phys. Rev. B} \textbf{54} 17954-61
	\bibitem{add89} Fujita M, Wakabayashi K, Nakada K and Kusakabe K 1996 Peculiar localized state at zigzag graphite edge \textit{J. Phys. Soc. Jpn.} \textbf{65} 1920-3
    \bibitem{add90} Su W P, Schrieffer J R and Heeger A J 1979 Solitons in polyacetylene \textit{Phys. Rev. Lett.} \textbf{42} 1698-701
	\bibitem{add91} Bistritzer R and MacDonald A H 2011 Moir\'{e} bands in twisted double-layer graphene \textit{Proc. Natl. Acad. Sci.} \textbf{108} 12233-7
	\bibitem{add92} Carr S, Fang S, Zhu Z and Kaxiras E 2019 Exact continuum model for low-energy electronic states of twisted bilayer graphene \textit{Phys. Rev. Res.} \textbf{1} 013001
	\bibitem{add93} Zhang Y-M, Antezza M and Wang J-S 2022 Controllable thermal radiation from twisted bilayer graphene \textit{Int. J. Heat Mass Transf.} \textbf{194} 123076
	\bibitem{add94} Freitag M, Chiu H-Y, Steiner M, Perebeinos V and Avouris P 2010 Thermal infrared emission from biased graphene \textit{Nat. Nanotechnol.} \textbf{5} 497-501
	\bibitem{add95} Lopes dos Santos J M B, Peres N M R and Castro Neto A H 2007 Graphene bilayer with a twist: Electronic structure \textit{Phys. Rev. Lett.} \textbf{99} 256802
	\bibitem{add96} Matsyshyn O, Song J C W, Villadiego I S and Shi L 2023 Fermi-Dirac staircase occupation of Floquet bands and current rectification inside the optical gap of metals: An exact approach \textit{Phys. Rev. B} \textbf{107} 195135
	\bibitem{add97} Rodriguez-Vega M, Vogl M and Fiete G A 2021 Low-frequency and Moir\'{e}-Floquet engineering: A review \textit{Ann. Phys.} \textbf{435} 168434
    \bibitem{add98} Morgado T A and Silveirinha M G 2017 Negative Landau damping in bilayer graphene \textit{Phys. Rev. Lett.} \textbf{119} 133901
    \bibitem{add99} Guo Y, Viloria M G, Messina R, Ben-Abdallah P, and Merabia S 2023 Atomistic modeling of extreme near-field heat transport across nanogaps between two polar dielectric materials \textit{Phys. Rev. B} \textbf{108} 085434
    \bibitem{add100}
    Viloria M G, Guo Y, Merabia S, Messina R, and Ben-Abdallah P 2023 Radiative heat exchange driven by acoustic vibration modes between two solids at the atomic scale \textit{Phys. Rev. B} \textbf{108} L201402
    \bibitem{add101}
    Rajabpour A, Hajj J E, Viloria M G, Messina R, Ben-Abdallah P, Guo Y, and Merabia S 2024 Extreme near-field heat transfer between silica surfaces \textit{Appl. Phys. Lett.} \textbf{124} 192203
    \bibitem{add102} Bhandari B, Erdman P A, Fazio R, Paladino E and Taddei F Thermal rectification through a nonlinear quantum resonator 2021 \textit{Phys. Rev. B} \textbf{103} 155434
    \bibitem{add103} Tang G and Wang J-S 2018 Heat transfer statistics in extreme-near-field radiation \textit{Phys. Rev. B} \textbf{98} 125401
\end{thebibliography}
\end{document}